\documentclass[aps,prl,twocolumn,amsmath,amssymb,superscriptaddress,floatfix]{revtex4-2}

\usepackage{mymacros}

\begin{document}
\title{Accurate and efficient Bloch-oscillation-enhanced atom interferometry}

\def\affitp  {\affiliation{Leibniz Universit\"at Hannover, Institut f\"ur Theoretische Physik, Appelstr. 2, D-30167 Hannover, Germany }}
\def\affiqo  {\affiliation{Leibniz Universit\"at Hannover, Institut f\"ur Quantenoptik, Welfengarten 1, D-30167 Hannover, Germany }}
\author{F.~Fitzek}
\email[email: ]{fitzek@iqo.uni-hannover.de}
\affitp
\affiqo
\author{J.-N.~Kirsten-Siem\ss}
\affitp
\affiqo
\author{E.~M.~Rasel}
\author{N.~Gaaloul}
\affiqo
\author{K.~Hammerer}
\email[email: ]{klemens.hammerer@itp.uni-hannover.de}
\affitp
\date{\today}

%%%%%%%%%%%%%%%%%%%%%%%%%%%%%%%%%%%%%%%%%%%%%%%%%%%%%%%%%%%%%%%%
%%%     Abstract
%%%%%%%%%%%%%%%%%%%%%%%%%%%%%%%%%%%%%%%%%%%%%%%%%%%%%%%%%%%%%%%%    
\begin{abstract}
Bloch oscillations of atoms in optical lattices are a powerful technique that can boost the sensitivity of atom interferometers to a wide range of signals by large momentum transfer. To leverage this method to its full potential, an accurate theoretical description of losses and phases is needed going beyond existing treatments. Here, we present a comprehensive theoretical framework for Bloch-oscillation-enhanced atom interferometry and verify its accuracy through comparison with an exact numerical solution of the Schrödinger equation. Our approach establishes design criteria to reach the fundamental efficiency and accuracy limits of large momentum transfer using Bloch oscillations. We compare these limits to the case of current state-of-the-art experiments and make projections for the next generation of quantum sensors.
\end{abstract}
\maketitle

%%%%%%%%%%%%%%%%%%%%%%%%%%%%%%%%%%%%%%%%%%%%%%%%%%%%%%%%%%%%%%%%
%%%     Section: Introduction
%%%%%%%%%%%%%%%%%%%%%%%%%%%%%%%%%%%%%%%%%%%%%%%%%%%%%%%%%%%%%%%%
\begin{figure}[t]
        \includegraphics[width=\columnwidth,keepaspectratio]{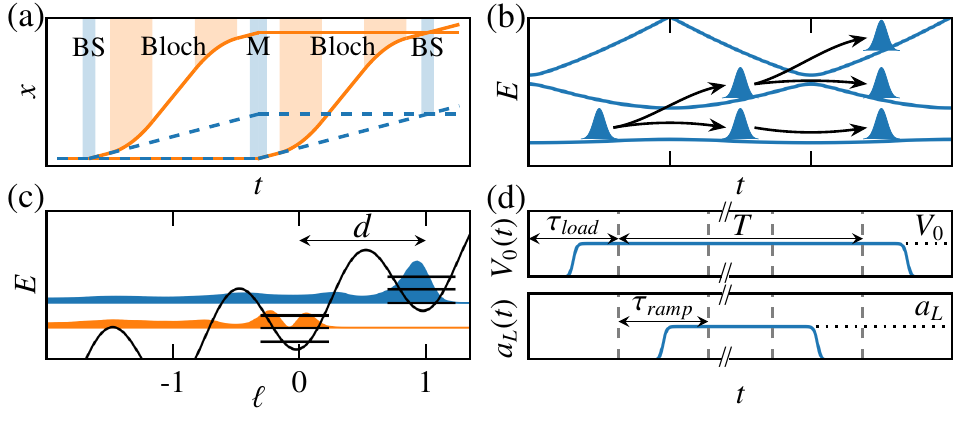}
        \caption{(a) Space-time diagram of a Mach-Zehnder LMT atom interferometer consisting of two beam splitter pulses (blue, BS) and one mirror pulse (blue, M). Four sequences of BOs (orange, Bloch) are used to sequentially accelerate and decelerate the arms of the interferometer. (b) Pictorial representation of LMT Bloch pulse in Bloch basis. (c) Pictorial representation of WS states $\ket*{\Psi_{\alpha=0,\,\ell=1}}$ and $\ket*{\Psi_{\alpha=1,\,\ell=0}}$ in tilted lattice potential [see Eq.~\eqref{eq:Hamiltonian}]. (d) Lattice depth $V_0(t)$ and acceleration $a_L(t)$ versus time $t$ for an LMT Bloch pulse, representing an acceleration or deceleration pulse, as depicted in (a). Our choice for $V_0(t)$ and $a_L(t)$ is a stretched pulse with sigmoidal rise and fall. 
        }\label{fig:figureOne}
\end{figure}

Large-momentum-transfer (LMT) techniques are essential tools to enhance the sensitivity of atom interferometers, which have emerged as versatile quantum sensors capable of highly accurate and precise measurements with numerous applications. These include the determination of fundamental constants like the fine-structure constant~\cite{Parker2018,Morel2020} and the gravitational constant~\cite{Rosi2014}, tests of general relativity~\cite{Dimopoulos2008a, Asenbaum2020} as well as applications in geophysics~\cite{Bongs2019} and inertial navigation~\cite{Geiger2011,Cheiney2018,Bongs2019,Geiger2020}. Bloch oscillations (BOs) of atoms in optical lattices~\cite{BenDahan1996, Wilkinson1996} are a frequently utilized LMT technique in state-of-the-art experiments to generate momentum changes of $100$ to $1000$ photon recoils to either measure the fine-structure constant~\cite{Parker2018,Morel2020}, hold atoms against gravity~\cite{Panda2022} or realize large spatial separations~\cite{Gebbe2021,Pagel2020}, as illustrated in  Fig.~\ref{fig:figureOne}(a). In these implementations, highly symmetric geometries or differential measurement techniques were required to suppress systematic phase uncertainties.  Operating atom interferometers beyond momentum changes of 1000 photon recoils is a critical requirement for the detection of gravitational waves in the mid-frequency band and the exploration of ultra-light dark matter and dark energy~\cite{Graham2013,Canuel2018,Canuel2020,Zhan2020,Badurina2020,Abe2021} as well as in advancing our understanding of the fine-structure constant measurement discrepancies~\cite{Morel2020}. For the continued progress of these fields it is essential to develop a model that accurately predicts losses, phases and phase uncertainties associated with BOs as an LMT technique. So far, control of the systematic phase uncertainty is lacking mainly due to the absence of a detailed theory of the phase build-up during a BO process.

The most common description of BOs is based on the adiabatic theorem using instantaneous Bloch states~\cite{Bloch1929,Zener1934}, as illustrated in Fig.~\ref{fig:figureOne}(b). Here, atoms are localized in the fundamental Bloch band and adiabatically follow the instantaneous eigenenergies for a non-vanishing acceleration of the optical lattice. The loss probability to neighbouring Bloch bands at avoided crossings is calculated using the Landau-Zener formula~\cite{Zener1934, SupplementFitzek2023}. It is known that this model fails to accurately describe BOs of atoms in optical lattices~\cite{Peik1997,Holthaus2000,Sias2007,Clade2017}, especially for deep lattice depths $V_0 \gtrsim 20\,E_r$. It is, however, in this regime that one would need to operate to realize sizable LMT processes. With growing lattice depth $V_0$ the Bloch bands become increasingly flat which violates the applicability of the Landau-Zener formula that relies on the existence of avoided crossings. As we will show, the predicted losses for a momentum transfer of $1000$ photon recoils based on the widely-used Landau-Zener formula differs by several orders of magnitude from the exact numerical solution. 
In this article, we develop a theoretical framework for BO-enhanced atom interferometers and establish design criteria for LMT Bloch pulses to reach their fundamental efficiency limit. We use our model to confirm the systematic limitations of current state-of-the-art experiments and regimes and verify its accuracy through comparison with an exact numerical integration of the Schrödinger equation.

%%%%%%%%%%%%%%%%%%%%%%%%%%%%%%%%%%%%%%%%%%%%%%%%%%%%%%%%%%%%%%%%
%%%     Section
%%%%%%%%%%%%%%%%%%%%%%%%%%%%%%%%%%%%%%%%%%%%%%%%%%%%%%%%%%%%%%%%
%\textcolor{black}{\textit{WS model.---}}
\begin{figure}[t]
        \includegraphics[width=\columnwidth,keepaspectratio]{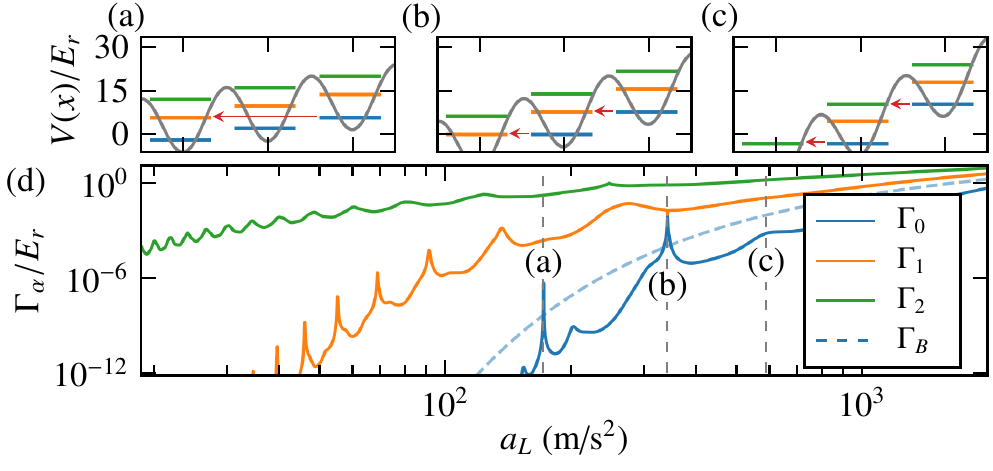}
        \caption{(a)-(c) Tilted potential [see Eq.~\eqref{eq:Hamiltonian}] for specific accelerations indicated in (d) and a lattice depth of $V_0=20\,E_r$, including corresponding WS energy levels. Red arrows represent process of resonant tunneling between different WS ladders. (d) Linewidth of WS ladders for $V_0=20\,E_r$ versus peak acceleration $a_L$. Solid lines show linewidths for the first three WS ladders $\alpha=0,1,2$, while the dashed line shows the linewidth for the fundamental WS ladder predicted by the Landau-Zener formula~\cite{SupplementFitzek2023}. %The true WS linewidths exhibit resonances due to the resonant tunneling processes shown in (a)-(c), in contrast to the monotonous exponential behaviour predicted by the Bloch model.
        }\label{fig:WSspectrum}
\end{figure}
We consider an atom with mass $m$ interacting with a pair of two counter-propagating light fields of adjustable intensity and frequency difference which gives rise to an optical lattice potential with lattice constant $d$ and wave number $k_L=\pi/d$. In a reference frame comoving with an accelerated optical lattice the Hamiltonian takes the form~\cite{SupplementFitzek2023}
\begin{align} \label{eq:Hamiltonian} 
        H(t)=\frac{\hat{p}^2}{2m}+V_0(t) \cos^2(k_L \hat{x})+ma_L(t)\hat{x}.
\end{align} 
Here, the time-dependent lattice depth can be expressed by the two-photon Rabi frequency $V_0(t)=2\hbar\Omega(t)$ and the acceleration of the optical lattice by the frequency difference of the counter-propagating light fields $a_L(t) = \frac{\pi}{k_L} \partial_t\Delta \nu(t)$~\cite{Peik1997}. An LMT Bloch pulse consists of three separate processes, as illustrated in Figs.~\ref{fig:figureOne}(d,e): First, during the loading time $\tau_{load}$ atoms are adiabatically loaded into the comoving optical lattice with peak lattice depth $V_0$. Second, atoms and lattice undergo an acceleration phase of duration $T$ characterized by a peak acceleration $a_L$ and a ramping time $\tau_{ramp}$. Finally, atoms are unloaded from the optical lattice and will have experienced a total momentum transfer of $2N\hbar k_L$ proportional to the number of BOs $N=ma_LT/2\hbar k_L$~\footnote{For non-vanishing ramping time $\tau_{ramp}$, $T$ is determined by integrating the acceleration curve $a_L(t)$ and solving for the final velocity $2N\hbar k_L/m$.}.

The basis of our theoretical description of the acceleration process will be so-called Wannier-Stark (WS) states $\ket*{\Psi_{\alpha,\ell}}$, which are the (quasi-)bound eigenstates of $H(t)$~\cite{Wannier1960, Nenciu1991}
\begin{align} \label{eq:WSstates}
        H(t)\ket*{\Psi_{\alpha,\ell}(t)}=\left(E_{\alpha,0}(t)+d\ell ma_L(t)-i\Gamma_{\alpha}(t)/2\right)\ket*{\Psi_{\alpha,\ell}(t)},
\end{align}
where $\ell$ denotes the lattice site quantum number and $\alpha$ the quantum number labelling the so-called $\alpha$th WS ladder. The eigenvalues are complex energies that depend on the instantaneous lattice depth $V_0(t)$ and acceleration $a_L(t)$. Their imaginary part $\Gamma_{\alpha}(t)$ represents a linewidth for the $\alpha$th WS ladder and describes tunneling losses~\cite{Nenciu1991}. As illustrated in Fig.~\ref{fig:figureOne}(c), atoms that are localized in the tilted potential [see Eq.~\eqref{eq:Hamiltonian}] can undergo tunneling events until they escape from the optical lattice. In this case, they behave like free particles represented by the infinite tail of WS states in opposite direction of motion of the optical lattice. To compute the complex WS energies in Eq.~\eqref{eq:WSstates} a numerical routine closely related to Floquet theory is required~\cite{Gluck2002, Clade2017}. We generalize this method to incorporate the treatment of adiabatic acceleration pulses, as depicted in Fig.~\ref{fig:figureOne}(d), by developing an energy sorting algorithm based on an analytical approximation for WS energies $E_{\alpha,l}$~\cite{SupplementFitzek2023}. Fig.~\ref{fig:WSspectrum}(d) displays that the resulting  WS linewidths $\Gamma_{\alpha}(t)$  depend non-trivially on $a_L(t)$, exhibiting tunneling resonances~\cite{Nenciu1991}. In contrast, the loss rate $\Gamma_B$ predicted on the basis of Bloch states and the Landau-Zener formula~\cite{SupplementFitzek2023} shows a monotonous behaviour and differs dramatically from the WS linewidth $\Gamma_0$. The position of the tunneling resonances can be explained by identifying crossings of the real part of the WS energies between different WS ladders, as shown in Figs.~\ref{fig:WSspectrum}(a)-(c). In the past, the properties and existence of the WS spectrum were controversially discussed~\cite{Wannier1960, Zak1968, Wannier1969, Zak1969} and only after decades of research efforts a rigorous justification of the spectral properties of $H(t)$ could be achieved~\cite{Nenciu1991}. Since then, WS states have been used to analyze experiments of cold atoms in accelerated optical lattices~\cite{Wilkinson1996, Gluck2002, Sias2007, Clade2017}. We advocate that a model based on WS states is the adequate picture to devise LMT Bloch pulses reaching their fundamental efficiency limit. 
\begin{figure}[t]
        \includegraphics[width=\columnwidth,keepaspectratio]{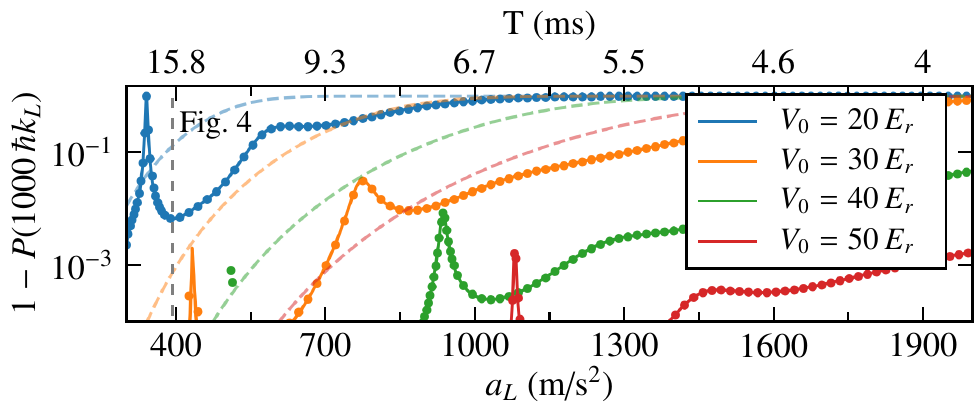}
        \caption{Losses for an adiabatic $1000\,\hbar k_L$ LMT Bloch pulse versus peak acceleration $a_L$ for a given lattice depth $V_0$ and ramping time of $\tau_{ramp}=1$ ms. The upper axis shows the corresponding acceleration time $T$ [see Fig.~\ref{fig:figureOne}(b)]. Solid lines represent losses based on the WS model [see Eq.~\eqref{eq:adiabaticWS}], while dots show the exact numerical solution. Dashed lines show the predicted losses based on the Landau-Zener formula~\cite{SupplementFitzek2023}. %All LMT Bloch pulses shown, use an acceleration ramp time of $\tau_{ramp}=1$ ms and sufficiently long loading times $\tau_{load}$ to populate only the fundamental Bloch band. The WS model and the numerical solution are in excellent agreement. 
        }\label{fig:Losses_over_aL}
\end{figure}

To realize optimal LMT Bloch pulses we propose adiabatic control of atoms in WS eigenstates. We first focus on the dynamics under the adiabatic approximation and then consider the control of non-adiabatic deviations. Fig.~\ref{fig:WSspectrum}(d) clearly demonstrates that due to the generally narrower linewidth corresponding to a longer lifetime, it is advantageous to drive LMT Bloch pulses in the fundamental WS ladder~\footnote{We discuss excited-band Bloch oscillations proposed by~\cite{McAlpine2020} in~\cite{SupplementFitzek2023}.}. This observation holds for arbitrary lattice depths~\cite{SupplementFitzek2023} and has direct implications for the process of loading atoms into the optical lattice, which we infer from the properties of WS states for vanishing accelerations. In this limit, we make use of the single-band approximation and neglect hopping processes to neighbouring lattices sites $\ell$, resulting in $\ket*{\Psi_{\alpha,\ell}}|_{a_L=0}=\ket*{w_{\alpha,\ell}}$, where $\ket*{w_{\alpha,\ell}}$ are Wannier states of the Bloch band $\alpha$ localized at lattice site $\ell$. Consequently, only atoms loaded into the fundamental Bloch band serve as a suitable initial condition for the acceleration phase to minimize tunneling losses. This is achieved for an atomic momentum distribution $\varphi (p)$ that vanishes outside the first Brillouin zone $[-\hbar k_L, \hbar k_L]$ and sufficiently long loading times $\tau_{load}$, as is common in state-of-the-art experiments~\cite{Parker2018,Morel2020,Gebbe2021,Pagel2020,Panda2022}.
If these requirements are violated atoms will populate excited WS ladders and thus experience increased tunneling losses, as seen in Fig.~\ref{fig:WSspectrum}(d). We apply the adiabatic approximation during the loading time $\tau_{load}$ and find $\ket*{\psi(\tau_{load})}=\sum_{\ell}g_{\ell}\ket*{w_{0,\ell}}$ with
\begin{align} \label{eq:adiabaticLoading}
    g_{\ell}=\int_{-\hbar k_L}^{\hbar k_L} \mathrm{d}p\, \varphi
    (p)\, e^{-i\int_0^{\tau_{load}}\mathrm{d}t' E_{0}(V_0(t'),\,p)/\hbar}e^{i p d \ell/\hbar}, 
\end{align}
where $E_{0}(V_0,p)$ are Bloch-eigenenergies of $H(t)$ for vanishing acceleration with a fixed lattice depth $V_0$ and quasi-momentum $p \in [-\hbar k_L, \hbar k_L]$. Based on Eq.~\eqref{eq:WSstates} we apply the adiabatic approximation during the acceleration time $T$ for Hamiltonians with discrete and non-degenerate complex spectrum~\cite{Mondragon1996, Keck2003, Ibanez2014}. That is, as long as there exists a finite complex energy gap between the fundamental and excited WS states and for sufficiently small changes of the acceleration $a_L(t)$, the time evolution of the system is governed by
\begin{align} \label{eq:adiabaticWS}
    \ket{\psi(t)}&=\sum_{\ell}e^{-i\int_0^t\mathrm{d}t'E_{0,0}(t')/\hbar}e^{-id\ell p_L(t)/\hbar}e^{-\int_0^t\mathrm{d}t'\Gamma_{0}(t')/2\hbar} \; g_{\ell} \ket*{\Psi_{0,\ell}(t)},
\end{align}
where the velocity of the optical lattice is given by $v_L(t)=p_L(t)/m=\int_0^t\mathrm{d}t'a_L(t')$. Eq.~\eqref{eq:adiabaticWS} describes phases and losses of the atomic wavepacket with a momentum distribution $\varphi (p)$ for $p \in [-\hbar k_L, \hbar k_L]$ undergoing an adiabatic LMT Bloch pulse.

The finite lifetime of WS states has important implications for LMT Bloch pulses. Indeed, Fig.~\ref{fig:Losses_over_aL} shows that for adiabatic LMT Bloch pulses the fundamental and dominant loss mechanism is given by tunneling losses. This is demonstrated by the excellent agreement between the adiabatic WS model [see Eq.~\eqref{eq:adiabaticWS}] and the exact numerical solution, where we solve the time-dependent Schrödinger equation for $H(t)$ on a position grid using the split-step Fourier method~\cite{Feit1982, Fitzek2020} with absorbing boundary conditions~\cite{Muga2004}. By means of our WS model, we identify combinations of optimal lattice depths and accelerations to minimize losses for LMT Bloch pulses. They are directly connected to the WS linewidth $\Gamma_0$ and the occurrence of resonant tunneling. In contrast, the losses predicted by the Landau-Zener formula~\cite{Bloch1929,Zener1934} deviate dramatically from the numerical solution for the entire range of parameters shown, with a general tendency towards greater deviations for increasing lattice depths.

We proceed to discuss non-adiabatic deviations of Eq.~\eqref{eq:adiabaticWS}. At tunneling resonances the complex energy gap can become very small and hence the tunneling probability from the fundamental to excited WS ladders increases when passing through them, leading to non-adiabatic losses. Once atoms are localized in an excited WS ladder $\alpha$, they are subject to increased tunneling losses quantified by $\Gamma_{\alpha}$, as shown in Fig.~\ref{fig:WSspectrum}(d), resulting in reduced fidelities for LMT Bloch pulses. Atoms that remain in excited WS ladders until the acceleration $a_L(t)$ is turned off are localized in excited Bloch bands, as evident from $\ket*{\Psi_{\alpha,\ell}}|_{a_L=0}=\ket*{w_{\alpha,\ell}}$. During the adiabatic unloading from the optical lattice, these atoms are mapped to momenta that differ, depending on the band number $\alpha$, by multiples of $\pm 2\hbar k_L$ from the target momentum $2N\hbar k_L$. Thus, in addition to the atomic trajectories shown in Fig.~\ref{fig:figureOne}(a) further trajectories are generated. For constant accelerations [see Fig.~\ref{fig:figureOne}(c)] the Hamiltonian $H(t)\equiv H$ will be time-independent and non-adiabatic losses cannot occur. This type of losses must be distinguished from tunneling losses quantified by $\Gamma_0$, which will always occur as long as there is a non-vanishing acceleration. 
%In particular, tunneling losses are the only fundamental loss channel during constant accelerations.

\begin{figure}[t]
        \includegraphics[width=\columnwidth,keepaspectratio]{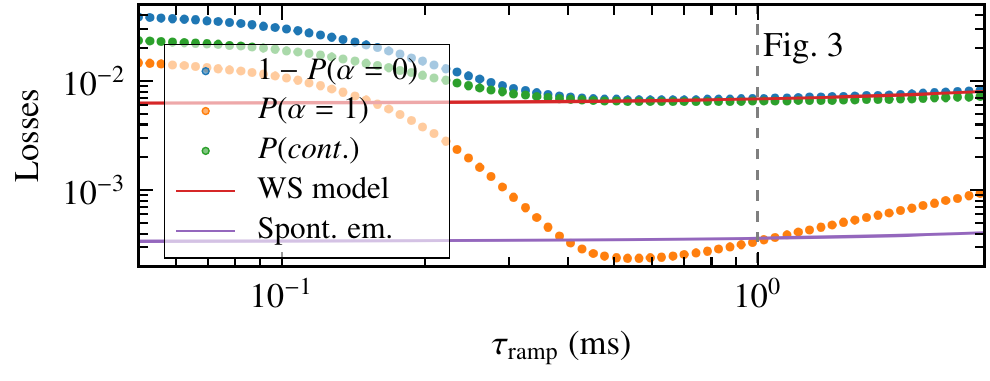}
        \caption{Losses for a $1000\,\hbar k_L$ LMT Bloch pulse versus acceleration ramp time $\tau_{ramp}$ for a peak lattice depth of
        $V_0=20\,E_r$. Red solid line represents losses based on the WS model, while dots show the results of exact numerical solutions, distinguishing between total losses from the WS ladder $\alpha=0$ (blue dots), losses to the WS ladder $\alpha=1$ (orange dots) and tunneling losses (green dots). Solid purple line shows spontaneous emission losses~\cite{SupplementFitzek2023}. The peak acceleration is optimally chosen at  $a_L=393.5$ $\mathrm{m/s}^2$, as determined from Fig.~\ref{fig:Losses_over_aL}. %We observe optimal acceleration ramp times of $\tau_{ramp}\gtrsim 0.5$ ms where the WS model and the numerical solution are in excellent agreement. These times correspond to adiabatic LMT Bloch pulses indicated by the significantly reduced non-adiabatic losses to the first excited WS ladder $P(\alpha=1)$.
        }\label{fig:Losses_over_tauAcc}
\end{figure}
The adiabaticity of an LMT Bloch pulse can be controlled by adjusting the acceleration ramp time $\tau_{ramp}$~\footnote{We discuss the lattice-shift method proposed by~\cite{Clade2017} in~\cite{SupplementFitzek2023}.}. In Fig.~\ref{fig:Losses_over_tauAcc}, we numerically analyze tunneling and non-adiabatic losses for an $1000\,\hbar k_L$ LMT Bloch pulse. For small times $\tau_{ramp}$ the flat-top pulse, as presented in Fig.~\ref{fig:figureOne}(d), resembles a box pulse causing a large fraction of atoms to tunnel to excited WS ladders during $\tau_{ramp}$. This results in an increased amount of tunneling losses with contributions from all WS ladders that were populated during the pulse, as shown in Fig.~\ref{fig:Losses_over_tauAcc}. For larger times $\tau_{ramp}$ non-adiabatic tunneling to excited WS ladders is significantly reduced and the total losses can be accurately computed using Eq.~\eqref{eq:adiabaticWS}. For increasingly larger times $\tau_{ramp}$ we observe a slight increase of non-adiabatic excitations due to the prolonged time atoms spend at the tunneling resonances. This leads to an optimal acceleration ramp time $\tau_{ramp}$ with minimal losses, which is accurately described by the tunneling losses of the fundamental WS ladder.

Apart from tunneling and non-adiabatic losses, atoms will also experience losses due to spontaneous emission. We estimate this additional loss channel based on a state-of-the-art laser system~\cite{Gebbe2021} for an optimal LMT Bloch pulse with minimal losses at peak lattice depth $V_0=20\,E_r$ and peak acceleration $a_L=393.5$ $\mathrm{m/s}^2$, as determined from Fig.~\ref{fig:Losses_over_aL}. In this setting, losses due to spontaneous emission are three times larger than tunneling losses~\cite{SupplementFitzek2023}, highlighting the need for more powerful laser systems to operate highly efficient LMT Bloch pulses. In comparison, with a recently developed enhanced laser system for atom interferometry~\cite{Kim2020} losses due to spontaneous emission are several orders of magnitude smaller than tunneling losses for moderate lattice depths $V_0 \lesssim 35 E_r$~\cite{SupplementFitzek2023}, as shown in Fig.~\ref{fig:Losses_over_tauAcc} for $V_0=20\,E_r$.

%%%%%%%%%%%%%%%%%%%%%%%%%%%%%%%%%%%%%%%%%%%%%%%%%%%%%%%%%%%%%%%%
%%%     Section
%%%%%%%%%%%%%%%%%%%%%%%%%%%%%%%%%%%%%%%%%%%%%%%%%%%%%%%%%%%%%%%%
\textcolor{black}{\textit{Phase uncertainties.---}}
\begin{figure}[t]
        \includegraphics[width=\columnwidth,keepaspectratio]{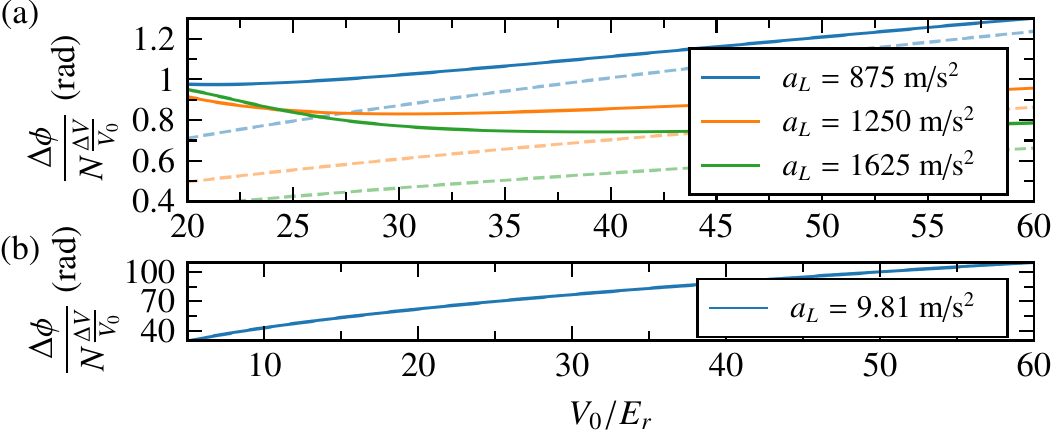}
        \caption{Phase uncertainty induced by lattice depth fluctuations versus peak lattice depth $V_0$ for different peak accelerations $a_L$. Solid lines represent the phase uncertainty based on the WS model, dashed lines show the phase uncertainty evaluated with the adiabatic Bloch model~\cite{SupplementFitzek2023} for (a) exemplary peak accelerations $a_L$ that allow for high-fidelity LMT Bloch pulses, as determined in Fig.~\ref{fig:Losses_over_aL} and (b) the local gravitational acceleration.
        }\label{fig:phase_uncertainty}
\end{figure}
The excellent agreement between the WS model and the exact numerical solution allows to accurately quantify systematic errors connected to adiabatic LMT Bloch pulses used in atom interferometers. A relevant contribution to the overall phase variance is induced by intensity fluctuations of the light pulse. Based on our model, the phase accrued in an adiabatic $2N\hbar k_L$ Bloch pulse in a single interferometer arm is given by $\phi=E_{0,0} N T_B/\hbar$, cf. Eq.~\eqref{eq:adiabaticWS}. A relative fluctuation $\Delta V$ in lattice depth between two arms, as illustrated in Fig.~\ref{fig:figureOne}(a), will therefore induce a variation of the relative phase 
\begin{align} \label{eq:phaseUncertainty}
        \frac{\Delta \phi} { N \;\dfrac{\Delta V}{V_0}}&=2\pi\left|\dfrac{\partial E_{0,0}}{\partial V_0} \right| \dfrac{V_0}{dma_L}.
\end{align}
Here, we refer the phase fluctuation to the number of BOs $N$ and the relative lattice depth fluctuation $\Delta V/V_0$. In Figs.~\ref{fig:phase_uncertainty}(a,b) we show phase changes for examplary accelerations, that allow for high-fidelity LMT Bloch pulses, displaying a general tendency towards larger phase uncertainties using smaller accelerations due to the prolonged time atoms spend in the optical lattice [see Eq.~\eqref{eq:phaseUncertainty}]. Based on Fig.~\ref{fig:phase_uncertainty}(a), we compute the required level of relative intensity stabilization to achieve a phase uncertainty of $\Delta \phi =1$ mrad, leading to $\Delta V / V_0 \simeq 10^{-6}$. This presents a significant challenge for utilizing LMT Bloch pulses in atom interferometers and highlights the advantage of symmetric atom interferometer geometries in reducing relative intensity fluctuations~\cite{Gebbe2021, Pagel2020}.

In the following, we use our model to explain the limitations of current state-of-the-art experiments~\cite{Morel2020, Panda2022a, Canuel2020}, where $\Delta V$ is due to a variance in the tilt angle $\Delta\theta$ of the beam axis, resulting in transversal displacements of the atomic clouds. As a first example, we analyze the fine-structure constant measurement presented in Morel \textit{et al.}~\cite{Morel2020, Morel2019}. This setup utilizes a Ramsey-Bord\'{e} geometry with $^{87}\mathrm{Rb}$ atoms, where both arms of the interferometer are accelerated simultaneously using one optical lattice with a momentum transfer of $1000\,\hbar k_L$. For small $\Delta\theta$ and a given separation between the interferometer arms $z$ this results in a relative lattice depth fluctuation of $\Delta V/ V_0 = \Delta \theta^2 z^2/ 2 w_0^2$, using the harmonic approximation for the Gaussian laser mode with waist $w_0$. To reach a relative uncertainty in $h/m$ at the level of $10^{-9}$ in one shot four central fringes need to be resolved with a phase uncertainty of $\Delta \phi = 1$ mrad, as reported in~\cite{Morel2019}. Using our model, we provide an upper bound for the relative intensity stabilization at $\Delta V/V_0\lesssim 1.51\times10^{-6}$ which is equivalent to a maximal variance of the tilt angle of $\Delta\theta \lesssim 16.5$ mrad. To improve the precision by one order of magnitude the variance needs to be bound by $\Delta\theta \lesssim 5.2$ mrad.

As a second example, we analyze the cavity experiment presented in Panda \textit{et al.}~\cite{Panda2022a}. In this setup, a spatially separated superposition of $^{133}\mathrm{Cs}$ atoms is adiabatically loaded into a vertically aligned optical lattice and held against gravity, cf. Fig.~\ref{fig:phase_uncertainty}(b). The record-breaking coherence time of one minute corresponds to $N\approx 92435$ BOs. Again, oscillatory tilts of the vertical cavity axis cause a relative lattice depth instability between the atomic clouds which amounts to $\Delta V / V_0 \simeq \Delta\theta  z /w_0$ in the case of a shallow optical lattice with $V_0=7\,E_r$~\cite{Panda2022a}. These oscillatory tilts lead to a loss of contrast, which are explained in~\cite{Panda2022a} on the basis of semi-classical Monte-Carlo simulations that model the transversal motion of atoms in the Gaussian beam. Alternatively, we infer from Eq.~\eqref{eq:phaseUncertainty} of our one-dimensional model a phase uncertainty of $\Delta \phi \approx 0.9$ rad for a holding time of one minute and a tilt uncertainty of $\Delta\theta=300$ $\mu$rad~\cite{Panda2022a}. This is consistent with the loss of contrast observed in the experiment. The temperature dependency of the observed contrast decay~\cite{Panda2022a} could be treated in our description by extending to a cylindrically symmetric description of the Gaussian beam.

%The experimental data shows a loss of contrast, which is caused by oscillatory tilts of the vertical cavity axis. Again, this results in a relative lattice depth instability between the atomic clouds which amounts to $\Delta V / V_0 \simeq \Delta\theta  z /w_0$ in the case of a shallow optical lattice with $V_0=7\,E_r$~\cite{Panda2022a}. 
%larger separations lead to larger phase uncertainties [see Eq.~\eqref{eq:phaseUncertainty}], which reduces the contrast. 

% 1 C wegen tilts 
% 2 Modelierung der transversalen Bewegung semi-classical Monte-Carlo 
% 3 We: abschätzen nur unter axialer bewegung schon da. 1D usw. 
% 4 Temperaturabh. spielt eine Rolle, mit unserem Model dafür aber at least cylindrical, i.e. 2D.

%The experimental data shows that larger spatial separations $z$ result in reduced contrasts, which can be explained with the help of a semi-classical Monte-Carlo simulation~\cite{Panda2022a}. Alternatively, we directly infer from our model that larger separations lead to larger phase uncertainties [see Eq.~\eqref{eq:phaseUncertainty}], which reduces the contrast. 
 
As a third example, we analyze a proposal for gravitational wave detection using atom interferometers presented in Canuel \textit{et al.}~\cite{Canuel2020, Canuel2020a}. The targeted baseline design foresees a double-loop geometry including LMT beam splitters based on Bragg diffraction~\cite{Martin1988, Giltner1995PRA} and BOs with $N=500$ and a phase resolution of $\Delta\phi=1\,\mu$rad. An optimal combination of lattice depth and acceleration for LMT Bloch pulses can be determined with the help of the WS model, as presented in Fig.~\ref{fig:Losses_over_aL}. For exemplary lattice depths and accelerations based on Fig.~\ref{fig:phase_uncertainty}(a) we quantify the necessary relative intensity stabilization between the two atomic trajectories in each interferometry arm to the level of $\Delta V/V_0\approx 10^{-9}.$ This result highlights the stringent requirements necessary for gravitational wave detection using LMT atom interferomters.

%%%%%%%%%%%%%%%%%%%%%%%%%%%%%%%%%%%%%%%%%%%%%%%%%%%%%%%%%%%%%%%%
%%% Conclusions
%%%%%%%%%%%%%%%%%%%%%%%%%%%%%%%%%%%%%%%%%%%%%%%%%%%%%%%%%%%%%%%% 
\textcolor{black}{\textit{Conclusions.---}} 
In summary, we have developed a model to design and evaluate BO-enhanced atom interferometers, providing insights into the fundamental loss processes as well as quantifying systematic errors. Using our framework, we show that there exist optimal combinations of ramping times $\tau_{ramp}$, peak lattice depths $V_0$ and accelerations $a_L$ to drive high-fidelity LMT Bloch pulses. To verify these predictions, we compare them with exact numerical solutions of the Schrödinger equation. Our theoretical description provides the basis for the consideration of other relevant systematic errors, such as phase noise or residual vibrations. These error sources can be assessed using the formalism of sensitivity functions~\cite{Cheinet2008}. Furthermore, we can consider the influence of spontaneous emission including the analysis of transversal effects~\cite{Pichler2010,Gluck2001}. In this context, our theoretical work presented in this article is the foundational piece for a comprehensive theoretical framework of LMT Bloch pulses for high-accuracy and high-precision atom interferometers.

%%%%%%%%%%%%%%%%%%%%%%%%%%%%%%%%%%%%%%%%%%%%%%%%%%%%%%%%%%%%%%%%
%%%     Section: Acknowledgements
%%%%%%%%%%%%%%%%%%%%%%%%%%%%%%%%%%%%%%%%%%%%%%%%%%%%%%%%%%%%%%%% 
\begin{acknowledgements}
After completion of this manuscript, we discussed with the group of S. Gupta~\footnote{T. Rahman et al., in preperation}, University of Washington, Seattle, about recent related experimental work. The results presented here were achieved by computations carried out on the cluster system at the Leibniz University of Hannover, Germany. This work was funded by the Deutsche Forschungsgemeinschaft (German Research Foundation) under Germany’s Excellence Strategy (EXC-2123 QuantumFrontiers Grants No. 390837967) and through CRC 1227 (DQ-mat) within Project No. A05, and the German Space Agency (DLR) with funds provided by the German Federal Ministry of Economic Affairs and Energy (German Federal Ministry of Education and Research (BMBF)) due to an enactment of the German Bundestag under Grants No. 50WM2250A (QUANTUS plus), No. 50WM2245A (CAL-II), No. 50WM2263A (CARIOQA-GE), No. 50WM2253A (AI-Quadrat) and No. 50NA2106 (QGYRO+).
\end{acknowledgements}

%%%%%%%%%%%%%%%%%%%%%%%%%%%%%%%%%%%%%%%%%%%%%%%%%%%%%%%%%%%%%%%%
%%%     Bibliography
%%%%%%%%%%%%%%%%%%%%%%%%%%%%%%%%%%%%%%%%%%%%%%%%%%%%%%%%%%%%%%%% 
\bibliographystyle{apsrev4-1_custom}
\bibliography{my_library}

\begin{thebibliography}{53}%
\makeatletter
\providecommand \@ifxundefined [1]{%
 \@ifx{#1\undefined}
}%
\providecommand \@ifnum [1]{%
 \ifnum #1\expandafter \@firstoftwo
 \else \expandafter \@secondoftwo
 \fi
}%
\providecommand \@ifx [1]{%
 \ifx #1\expandafter \@firstoftwo
 \else \expandafter \@secondoftwo
 \fi
}%
\providecommand \natexlab [1]{#1}%
\providecommand \enquote  [1]{``#1''}%
\providecommand \bibnamefont  [1]{#1}%
\providecommand \bibfnamefont [1]{#1}%
\providecommand \citenamefont [1]{#1}%
\providecommand \href@noop [0]{\@secondoftwo}%
\providecommand \href [0]{\begingroup \@sanitize@url \@href}%
\providecommand \@href[1]{\@@startlink{#1}\@@href}%
\providecommand \@@href[1]{\endgroup#1\@@endlink}%
\providecommand \@sanitize@url [0]{\catcode `\\12\catcode `\$12\catcode
  `\&12\catcode `\#12\catcode `\^12\catcode `\_12\catcode `\%12\relax}%
\providecommand \@@startlink[1]{}%
\providecommand \@@endlink[0]{}%
\providecommand \url  [0]{\begingroup\@sanitize@url \@url }%
\providecommand \@url [1]{\endgroup\@href {#1}{\urlprefix }}%
\providecommand \urlprefix  [0]{URL }%
\providecommand \Eprint [0]{\href }%
\providecommand \doibase [0]{http://dx.doi.org/}%
\providecommand \selectlanguage [0]{\@gobble}%
\providecommand \bibinfo  [0]{\@secondoftwo}%
\providecommand \bibfield  [0]{\@secondoftwo}%
\providecommand \translation [1]{[#1]}%
\providecommand \BibitemOpen [0]{}%
\providecommand \bibitemStop [0]{}%
\providecommand \bibitemNoStop [0]{.\EOS\space}%
\providecommand \EOS [0]{\spacefactor3000\relax}%
\providecommand \BibitemShut  [1]{\csname bibitem#1\endcsname}%
\let\auto@bib@innerbib\@empty
%</preamble>
\bibitem [{\citenamefont {Parker}\ \emph {et~al.}(2018)\citenamefont {Parker},
  \citenamefont {Yu}, \citenamefont {Zhong}, \citenamefont {Estey},\ and\
  \citenamefont {M{\"u}ller}}]{Parker2018}%
  \BibitemOpen
  \bibfield  {author} {\bibinfo {author} {\bibfnamefont {R.~H.}\ \bibnamefont
  {Parker}}, \bibinfo {author} {\bibfnamefont {C.}~\bibnamefont {Yu}}, \bibinfo
  {author} {\bibfnamefont {W.}~\bibnamefont {Zhong}}, \bibinfo {author}
  {\bibfnamefont {B.}~\bibnamefont {Estey}}, \ and\ \bibinfo {author}
  {\bibfnamefont {H.}~\bibnamefont {M{\"u}ller}},\ }\bibfield  {title} {\emph
  {\bibinfo {title} {Measurement of the fine-structure constant as a test of
  the {{Standard Model}}},\ }}\href {\doibase10.1126/science.aap7706}
  {\bibfield  {journal} {\bibinfo  {journal} {Science (New York, N.Y.)}\ }
  (\bibinfo {year} {2018}),\ 10.1126/science.aap7706}\BibitemShut {NoStop}%
\bibitem [{\citenamefont {Morel}\ \emph {et~al.}(2020)\citenamefont {Morel},
  \citenamefont {Yao}, \citenamefont {Clad{\'e}},\ and\ \citenamefont
  {{Guellati-Kh{\'e}lifa}}}]{Morel2020}%
  \BibitemOpen
  \bibfield  {author} {\bibinfo {author} {\bibfnamefont {L.}~\bibnamefont
  {Morel}}, \bibinfo {author} {\bibfnamefont {Z.}~\bibnamefont {Yao}}, \bibinfo
  {author} {\bibfnamefont {P.}~\bibnamefont {Clad{\'e}}}, \ and\ \bibinfo
  {author} {\bibfnamefont {S.}~\bibnamefont {{Guellati-Kh{\'e}lifa}}},\
  }\bibfield  {title} {\emph {\bibinfo {title} {Determination of the
  fine-structure constant with an accuracy of 81 parts per trillion},\ }}\href
  {\doibase10.1038/s41586-020-2964-7} {\bibfield  {journal} {\bibinfo
  {journal} {Nature}\ }\textbf {\bibinfo {volume} {588}},\ \bibinfo {pages}
  {61} (\bibinfo {year} {2020})}\BibitemShut {NoStop}%
\bibitem [{\citenamefont {Rosi}\ \emph {et~al.}(2014)\citenamefont {Rosi},
  \citenamefont {Sorrentino}, \citenamefont {Cacciapuoti}, \citenamefont
  {Prevedelli},\ and\ \citenamefont {Tino}}]{Rosi2014}%
  \BibitemOpen
  \bibfield  {author} {\bibinfo {author} {\bibfnamefont {G.}~\bibnamefont
  {Rosi}}, \bibinfo {author} {\bibfnamefont {F.}~\bibnamefont {Sorrentino}},
  \bibinfo {author} {\bibfnamefont {L.}~\bibnamefont {Cacciapuoti}}, \bibinfo
  {author} {\bibfnamefont {M.}~\bibnamefont {Prevedelli}}, \ and\ \bibinfo
  {author} {\bibfnamefont {G.~M.}\ \bibnamefont {Tino}},\ }\bibfield  {title}
  {\emph {\bibinfo {title} {Precision measurement of the {{Newtonian}}
  gravitational constant using cold atoms},\ }}\href
  {\doibase10.1038/nature13433} {\bibfield  {journal} {\bibinfo  {journal}
  {Nature}\ }\textbf {\bibinfo {volume} {510}},\ \bibinfo {pages} {518}
  (\bibinfo {year} {2014})}\BibitemShut {NoStop}%
\bibitem [{\citenamefont {Dimopoulos}\ \emph {et~al.}(2008)\citenamefont
  {Dimopoulos}, \citenamefont {Graham}, \citenamefont {Hogan},\ and\
  \citenamefont {Kasevich}}]{Dimopoulos2008a}%
  \BibitemOpen
  \bibfield  {author} {\bibinfo {author} {\bibfnamefont {S.}~\bibnamefont
  {Dimopoulos}}, \bibinfo {author} {\bibfnamefont {P.~W.}\ \bibnamefont
  {Graham}}, \bibinfo {author} {\bibfnamefont {J.~M.}\ \bibnamefont {Hogan}}, \
  and\ \bibinfo {author} {\bibfnamefont {M.~A.}\ \bibnamefont {Kasevich}},\
  }\bibfield  {title} {\emph {\bibinfo {title} {General relativistic effects in
  atom interferometry},\ }}\href {\doibase10.1103/PhysRevD.78.042003}
  {\bibfield  {journal} {\bibinfo  {journal} {Physical Review D}\ }\textbf
  {\bibinfo {volume} {78}},\ \bibinfo {pages} {042003} (\bibinfo {year}
  {2008})}\BibitemShut {NoStop}%
\bibitem [{\citenamefont {Asenbaum}\ \emph {et~al.}(2020)\citenamefont
  {Asenbaum}, \citenamefont {Overstreet}, \citenamefont {Kim}, \citenamefont
  {Curti},\ and\ \citenamefont {Kasevich}}]{Asenbaum2020}%
  \BibitemOpen
  \bibfield  {author} {\bibinfo {author} {\bibfnamefont {P.}~\bibnamefont
  {Asenbaum}}, \bibinfo {author} {\bibfnamefont {C.}~\bibnamefont
  {Overstreet}}, \bibinfo {author} {\bibfnamefont {M.}~\bibnamefont {Kim}},
  \bibinfo {author} {\bibfnamefont {J.}~\bibnamefont {Curti}}, \ and\ \bibinfo
  {author} {\bibfnamefont {M.~A.}\ \bibnamefont {Kasevich}},\ }\bibfield
  {title} {\emph {\bibinfo {title} {Atom-{{Interferometric Test}} of the
  {{Equivalence Principle}} at the $10^{-12}$ {{Level}}},\ }}\href
  {\doibase10.1103/PhysRevLett.125.191101} {\bibfield  {journal} {\bibinfo
  {journal} {Physical Review Letters}\ }\textbf {\bibinfo {volume} {125}},\
  \bibinfo {pages} {191101} (\bibinfo {year} {2020})}\BibitemShut {NoStop}%
\bibitem [{\citenamefont {Bongs}\ \emph {et~al.}(2019)\citenamefont {Bongs},
  \citenamefont {Holynski}, \citenamefont {Vovrosh}, \citenamefont {Bouyer},
  \citenamefont {Condon}, \citenamefont {Rasel}, \citenamefont {Schubert},
  \citenamefont {Schleich},\ and\ \citenamefont {Roura}}]{Bongs2019}%
  \BibitemOpen
  \bibfield  {author} {\bibinfo {author} {\bibfnamefont {K.}~\bibnamefont
  {Bongs}}, \bibinfo {author} {\bibfnamefont {M.}~\bibnamefont {Holynski}},
  \bibinfo {author} {\bibfnamefont {J.}~\bibnamefont {Vovrosh}}, \bibinfo
  {author} {\bibfnamefont {P.}~\bibnamefont {Bouyer}}, \bibinfo {author}
  {\bibfnamefont {G.}~\bibnamefont {Condon}}, \bibinfo {author} {\bibfnamefont
  {E.}~\bibnamefont {Rasel}}, \bibinfo {author} {\bibfnamefont
  {C.}~\bibnamefont {Schubert}}, \bibinfo {author} {\bibfnamefont {W.~P.}\
  \bibnamefont {Schleich}}, \ and\ \bibinfo {author} {\bibfnamefont
  {A.}~\bibnamefont {Roura}},\ }\bibfield  {title} {\emph {\bibinfo {title}
  {Taking atom interferometric quantum sensors from the laboratory to
  real-world applications},\ }}\href {\doibase10.1038/s42254-019-0117-4}
  {\bibfield  {journal} {\bibinfo  {journal} {Nature Reviews Physics}\ }\textbf
  {\bibinfo {volume} {1}},\ \bibinfo {pages} {731} (\bibinfo {year}
  {2019})}\BibitemShut {NoStop}%
\bibitem [{\citenamefont {Geiger}\ \emph {et~al.}(2011)\citenamefont {Geiger}
  \emph {et~al.}}]{Geiger2011}%
  \BibitemOpen
  \bibfield  {author} {\bibinfo {author} {\bibfnamefont {R.}~\bibnamefont
  {Geiger}} \emph {et~al.},\ }\bibfield  {title} {\emph {\bibinfo {title}
  {Detecting inertial effects with airborne matter-wave interferometry},\
  }}\href {\doibase10.1038/ncomms1479} {\bibfield  {journal} {\bibinfo
  {journal} {Nature Communications}\ }\textbf {\bibinfo {volume} {2}},\
  \bibinfo {pages} {474} (\bibinfo {year} {2011})}\BibitemShut {NoStop}%
\bibitem [{\citenamefont {Cheiney}\ \emph {et~al.}(2018)\citenamefont
  {Cheiney}, \citenamefont {Fouch{\'e}}, \citenamefont {Templier},
  \citenamefont {Napolitano}, \citenamefont {Battelier}, \citenamefont
  {Bouyer},\ and\ \citenamefont {Barrett}}]{Cheiney2018}%
  \BibitemOpen
  \bibfield  {author} {\bibinfo {author} {\bibfnamefont {P.}~\bibnamefont
  {Cheiney}}, \bibinfo {author} {\bibfnamefont {L.}~\bibnamefont {Fouch{\'e}}},
  \bibinfo {author} {\bibfnamefont {S.}~\bibnamefont {Templier}}, \bibinfo
  {author} {\bibfnamefont {F.}~\bibnamefont {Napolitano}}, \bibinfo {author}
  {\bibfnamefont {B.}~\bibnamefont {Battelier}}, \bibinfo {author}
  {\bibfnamefont {P.}~\bibnamefont {Bouyer}}, \ and\ \bibinfo {author}
  {\bibfnamefont {B.}~\bibnamefont {Barrett}},\ }\bibfield  {title} {\emph
  {\bibinfo {title} {Navigation-{{Compatible Hybrid Quantum Accelerometer
  Using}} a {{Kalman Filter}}},\ }}\href
  {\doibase10.1103/PhysRevApplied.10.034030} {\bibfield  {journal} {\bibinfo
  {journal} {Physical Review Applied}\ }\textbf {\bibinfo {volume} {10}},\
  \bibinfo {pages} {034030} (\bibinfo {year} {2018})}\BibitemShut {NoStop}%
\bibitem [{\citenamefont {Geiger}\ \emph {et~al.}(2020)\citenamefont {Geiger},
  \citenamefont {Landragin}, \citenamefont {Merlet},\ and\ \citenamefont
  {Pereira Dos~Santos}}]{Geiger2020}%
  \BibitemOpen
  \bibfield  {author} {\bibinfo {author} {\bibfnamefont {R.}~\bibnamefont
  {Geiger}}, \bibinfo {author} {\bibfnamefont {A.}~\bibnamefont {Landragin}},
  \bibinfo {author} {\bibfnamefont {S.}~\bibnamefont {Merlet}}, \ and\ \bibinfo
  {author} {\bibfnamefont {F.}~\bibnamefont {Pereira Dos~Santos}},\ }\bibfield
  {title} {\emph {\bibinfo {title} {High-accuracy inertial measurements with
  cold-atom sensors},\ }}\href {\doibase10.1116/5.0009093} {\bibfield
  {journal} {\bibinfo  {journal} {AVS Quantum Science}\ }\textbf {\bibinfo
  {volume} {2}},\ \bibinfo {pages} {024702} (\bibinfo {year}
  {2020})}\BibitemShut {NoStop}%
\bibitem [{\citenamefont {Ben~Dahan}\ \emph {et~al.}(1996)\citenamefont
  {Ben~Dahan}, \citenamefont {Peik}, \citenamefont {Reichel}, \citenamefont
  {Castin},\ and\ \citenamefont {Salomon}}]{BenDahan1996}%
  \BibitemOpen
  \bibfield  {author} {\bibinfo {author} {\bibfnamefont {M.}~\bibnamefont
  {Ben~Dahan}}, \bibinfo {author} {\bibfnamefont {E.}~\bibnamefont {Peik}},
  \bibinfo {author} {\bibfnamefont {J.}~\bibnamefont {Reichel}}, \bibinfo
  {author} {\bibfnamefont {Y.}~\bibnamefont {Castin}}, \ and\ \bibinfo {author}
  {\bibfnamefont {C.}~\bibnamefont {Salomon}},\ }\bibfield  {title} {\emph
  {\bibinfo {title} {Bloch {{Oscillations}} of {{Atoms}} in an {{Optical
  Potential}}},\ }}\href {\doibase10.1103/PhysRevLett.76.4508} {\bibfield
  {journal} {\bibinfo  {journal} {Physical Review Letters}\ }\textbf {\bibinfo
  {volume} {76}},\ \bibinfo {pages} {4508} (\bibinfo {year}
  {1996})}\BibitemShut {NoStop}%
\bibitem [{\citenamefont {Wilkinson}\ \emph {et~al.}(1996)\citenamefont
  {Wilkinson}, \citenamefont {Bharucha}, \citenamefont {Madison}, \citenamefont
  {Niu},\ and\ \citenamefont {Raizen}}]{Wilkinson1996}%
  \BibitemOpen
  \bibfield  {author} {\bibinfo {author} {\bibfnamefont {S.~R.}\ \bibnamefont
  {Wilkinson}}, \bibinfo {author} {\bibfnamefont {C.~F.}\ \bibnamefont
  {Bharucha}}, \bibinfo {author} {\bibfnamefont {K.~W.}\ \bibnamefont
  {Madison}}, \bibinfo {author} {\bibfnamefont {Q.}~\bibnamefont {Niu}}, \ and\
  \bibinfo {author} {\bibfnamefont {M.~G.}\ \bibnamefont {Raizen}},\ }\bibfield
   {title} {\emph {\bibinfo {title} {Observation of {{Atomic Wannier-Stark
  Ladders}} in an {{Accelerating Optical Potential}}},\ }}\href
  {\doibase10.1103/PhysRevLett.76.4512} {\bibfield  {journal} {\bibinfo
  {journal} {Physical Review Letters}\ }\textbf {\bibinfo {volume} {76}},\
  \bibinfo {pages} {4512} (\bibinfo {year} {1996})}\BibitemShut {NoStop}%
\bibitem [{\citenamefont {Panda}\ \emph
  {et~al.}(2022{\natexlab{a}})\citenamefont {Panda}, \citenamefont {Tao},
  \citenamefont {Egelhoff}, \citenamefont {Ceja}, \citenamefont {Xu},\ and\
  \citenamefont {M{\"u}ller}}]{Panda2022}%
  \BibitemOpen
  \bibfield  {author} {\bibinfo {author} {\bibfnamefont {C.~D.}\ \bibnamefont
  {Panda}}, \bibinfo {author} {\bibfnamefont {M.}~\bibnamefont {Tao}}, \bibinfo
  {author} {\bibfnamefont {J.}~\bibnamefont {Egelhoff}}, \bibinfo {author}
  {\bibfnamefont {M.}~\bibnamefont {Ceja}}, \bibinfo {author} {\bibfnamefont
  {V.}~\bibnamefont {Xu}}, \ and\ \bibinfo {author} {\bibfnamefont
  {H.}~\bibnamefont {M{\"u}ller}},\ }\href {\doibase10.48550/arXiv.2210.07289}
  {\bibinfo {title} {Quantum metrology by one-minute interrogation of a
  coherent atomic spatial superposition},\ } (\bibinfo {year}
  {2022}{\natexlab{a}}),\ \Eprint {http://arxiv.org/abs/2210.07289}
  {arxiv:2210.07289 [gr-qc, physics:physics, physics:quant-ph]} \BibitemShut
  {NoStop}%
\bibitem [{\citenamefont {Gebbe}\ \emph {et~al.}(2021)\citenamefont {Gebbe}
  \emph {et~al.}}]{Gebbe2021}%
  \BibitemOpen
  \bibfield  {author} {\bibinfo {author} {\bibfnamefont {M.}~\bibnamefont
  {Gebbe}} \emph {et~al.},\ }\bibfield  {title} {\emph {\bibinfo {title}
  {Twin-lattice atom interferometry},\ }}\href
  {\doibase10.1038/s41467-021-22823-8} {\bibfield  {journal} {\bibinfo
  {journal} {Nature Communications}\ }\textbf {\bibinfo {volume} {12}},\
  \bibinfo {pages} {2544} (\bibinfo {year} {2021})}\BibitemShut {NoStop}%
\bibitem [{\citenamefont {Pagel}\ \emph {et~al.}(2020)\citenamefont {Pagel},
  \citenamefont {Zhong}, \citenamefont {Parker}, \citenamefont {Olund},
  \citenamefont {Yao},\ and\ \citenamefont {M{\"u}ller}}]{Pagel2020}%
  \BibitemOpen
  \bibfield  {author} {\bibinfo {author} {\bibfnamefont {Z.}~\bibnamefont
  {Pagel}}, \bibinfo {author} {\bibfnamefont {W.}~\bibnamefont {Zhong}},
  \bibinfo {author} {\bibfnamefont {R.~H.}\ \bibnamefont {Parker}}, \bibinfo
  {author} {\bibfnamefont {C.~T.}\ \bibnamefont {Olund}}, \bibinfo {author}
  {\bibfnamefont {N.~Y.}\ \bibnamefont {Yao}}, \ and\ \bibinfo {author}
  {\bibfnamefont {H.}~\bibnamefont {M{\"u}ller}},\ }\bibfield  {title} {\emph
  {\bibinfo {title} {Symmetric {{Bloch}} oscillations of matter waves},\
  }}\href {\doibase10.1103/PhysRevA.102.053312} {\bibfield  {journal} {\bibinfo
   {journal} {Physical Review A}\ }\textbf {\bibinfo {volume} {102}},\ \bibinfo
  {pages} {053312} (\bibinfo {year} {2020})}\BibitemShut {NoStop}%
\bibitem [{\citenamefont {Graham}\ \emph {et~al.}(2013)\citenamefont {Graham},
  \citenamefont {Hogan}, \citenamefont {Kasevich},\ and\ \citenamefont
  {Rajendran}}]{Graham2013}%
  \BibitemOpen
  \bibfield  {author} {\bibinfo {author} {\bibfnamefont {P.~W.}\ \bibnamefont
  {Graham}}, \bibinfo {author} {\bibfnamefont {J.~M.}\ \bibnamefont {Hogan}},
  \bibinfo {author} {\bibfnamefont {M.~A.}\ \bibnamefont {Kasevich}}, \ and\
  \bibinfo {author} {\bibfnamefont {S.}~\bibnamefont {Rajendran}},\ }\bibfield
  {title} {\emph {\bibinfo {title} {New {{Method}} for {{Gravitational Wave
  Detection}} with {{Atomic Sensors}}},\ }}\href
  {\doibase10.1103/PhysRevLett.110.171102} {\bibfield  {journal} {\bibinfo
  {journal} {Physical Review Letters}\ }\textbf {\bibinfo {volume} {110}},\
  \bibinfo {pages} {171102} (\bibinfo {year} {2013})}\BibitemShut {NoStop}%
\bibitem [{\citenamefont {Canuel}\ \emph {et~al.}(2018)\citenamefont {Canuel},
  \citenamefont {Bertoldi}, \citenamefont {Amand} \emph {et~al.}}]{Canuel2018}%
  \BibitemOpen
  \bibfield  {author} {\bibinfo {author} {\bibfnamefont {B.}~\bibnamefont
  {Canuel}}, \bibinfo {author} {\bibfnamefont {A.}~\bibnamefont {Bertoldi}},
  \bibinfo {author} {\bibfnamefont {L.}~\bibnamefont {Amand}},  \emph
  {et~al.},\ }\bibfield  {title} {\emph {\bibinfo {title} {Exploring gravity
  with the {{MIGA}} large scale atom interferometer},\ }}\href
  {\doibase10.1038/s41598-018-32165-z} {\bibfield  {journal} {\bibinfo
  {journal} {Scientific Reports}\ }\textbf {\bibinfo {volume} {8}},\ \bibinfo
  {pages} {14064} (\bibinfo {year} {2018})}\BibitemShut {NoStop}%
\bibitem [{\citenamefont {Canuel}\ \emph
  {et~al.}(2020{\natexlab{a}})\citenamefont {Canuel} \emph
  {et~al.}}]{Canuel2020}%
  \BibitemOpen
  \bibfield  {author} {\bibinfo {author} {\bibfnamefont {B.}~\bibnamefont
  {Canuel}} \emph {et~al.},\ }\bibfield  {title} {\emph {\bibinfo {title}
  {{{ELGAR}}\textemdash a {{European Laboratory}} for {{Gravitation}} and
  {{Atom-interferometric Research}}},\ }}\href
  {\doibase10.1088/1361-6382/aba80e} {\bibfield  {journal} {\bibinfo  {journal}
  {Classical and Quantum Gravity}\ }\textbf {\bibinfo {volume} {37}},\ \bibinfo
  {pages} {225017} (\bibinfo {year} {2020}{\natexlab{a}})}\BibitemShut
  {NoStop}%
\bibitem [{\citenamefont {Zhan}\ \emph {et~al.}(2020)\citenamefont {Zhan} \emph
  {et~al.}}]{Zhan2020}%
  \BibitemOpen
  \bibfield  {author} {\bibinfo {author} {\bibfnamefont {M.-S.}\ \bibnamefont
  {Zhan}} \emph {et~al.},\ }\bibfield  {title} {\emph {\bibinfo {title}
  {{{ZAIGA}}: {{Zhaoshan}} long-baseline atom interferometer gravitation
  antenna},\ }}\href {\doibase10.1142/S0218271819400054} {\bibfield  {journal}
  {\bibinfo  {journal} {International Journal of Modern Physics D}\ }\textbf
  {\bibinfo {volume} {29}},\ \bibinfo {pages} {1940005} (\bibinfo {year}
  {2020})}\BibitemShut {NoStop}%
\bibitem [{\citenamefont {Badurina}\ \emph {et~al.}(2020)\citenamefont
  {Badurina} \emph {et~al.}}]{Badurina2020}%
  \BibitemOpen
  \bibfield  {author} {\bibinfo {author} {\bibfnamefont {L.}~\bibnamefont
  {Badurina}} \emph {et~al.},\ }\bibfield  {title} {\emph {\bibinfo {title}
  {{{AION}}: {{An}} atom interferometer observatory and network},\ }}\href
  {\doibase10.1088/1475-7516/2020/05/011} {\bibfield  {journal} {\bibinfo
  {journal} {Journal of Cosmology and Astroparticle Physics}\ }\textbf
  {\bibinfo {volume} {2020}},\ \bibinfo {pages} {011} (\bibinfo {year}
  {2020})}\BibitemShut {NoStop}%
\bibitem [{\citenamefont {Abe}\ \emph {et~al.}(2021)\citenamefont {Abe} \emph
  {et~al.}}]{Abe2021}%
  \BibitemOpen
  \bibfield  {author} {\bibinfo {author} {\bibfnamefont {M.}~\bibnamefont
  {Abe}} \emph {et~al.},\ }\bibfield  {title} {\emph {\bibinfo {title}
  {Matter-wave {{Atomic Gradiometer Interferometric Sensor}} ({{MAGIS-100}})},\
  }}\href {\doibase10.1088/2058-9565/abf719} {\bibfield  {journal} {\bibinfo
  {journal} {Quantum Science and Technology}\ }\textbf {\bibinfo {volume}
  {6}},\ \bibinfo {pages} {044003} (\bibinfo {year} {2021})}\BibitemShut
  {NoStop}%
\bibitem [{\citenamefont {Bloch}(1929)}]{Bloch1929}%
  \BibitemOpen
  \bibfield  {author} {\bibinfo {author} {\bibfnamefont {F.}~\bibnamefont
  {Bloch}},\ }\bibfield  {title} {\emph {\bibinfo {title} {{\"Uber die
  Quantenmechanik der Elektronen in Kristallgittern}},\ }}\href
  {\doibase10.1007/BF01339455} {\bibfield  {journal} {\bibinfo  {journal}
  {Zeitschrift f\"ur Physik}\ }\textbf {\bibinfo {volume} {52}},\ \bibinfo
  {pages} {555} (\bibinfo {year} {1929})}\BibitemShut {NoStop}%
\bibitem [{\citenamefont {Zener}\ and\ \citenamefont
  {Fowler}(1934)}]{Zener1934}%
  \BibitemOpen
  \bibfield  {author} {\bibinfo {author} {\bibfnamefont {C.}~\bibnamefont
  {Zener}}\ and\ \bibinfo {author} {\bibfnamefont {R.~H.}\ \bibnamefont
  {Fowler}},\ }\bibfield  {title} {\emph {\bibinfo {title} {A theory of the
  electrical breakdown of solid dielectrics},\ }}\href
  {\doibase10.1098/rspa.1934.0116} {\bibfield  {journal} {\bibinfo  {journal}
  {Proceedings of the Royal Society of London. Series A, Containing Papers of a
  Mathematical and Physical Character}\ }\textbf {\bibinfo {volume} {145}},\
  \bibinfo {pages} {523} (\bibinfo {year} {1934})}\BibitemShut {NoStop}%
\bibitem [{Sup()}]{SupplementFitzek2023}%
  \BibitemOpen
  \href@noop {} {\bibinfo {title} {See {{Supplemental Material}} at [{{URL}}
  will be inserted by publisher] for ...}\ }\BibitemShut {NoStop}%
\bibitem [{\citenamefont {Peik}\ \emph {et~al.}(1997)\citenamefont {Peik},
  \citenamefont {Ben~Dahan}, \citenamefont {Bouchoule}, \citenamefont
  {Castin},\ and\ \citenamefont {Salomon}}]{Peik1997}%
  \BibitemOpen
  \bibfield  {author} {\bibinfo {author} {\bibfnamefont {E.}~\bibnamefont
  {Peik}}, \bibinfo {author} {\bibfnamefont {M.}~\bibnamefont {Ben~Dahan}},
  \bibinfo {author} {\bibfnamefont {I.}~\bibnamefont {Bouchoule}}, \bibinfo
  {author} {\bibfnamefont {Y.}~\bibnamefont {Castin}}, \ and\ \bibinfo {author}
  {\bibfnamefont {C.}~\bibnamefont {Salomon}},\ }\bibfield  {title} {\emph
  {\bibinfo {title} {Bloch oscillations of atoms, adiabatic rapid passage, and
  monokinetic atomic beams},\ }}\href {\doibase10.1103/PhysRevA.55.2989}
  {\bibfield  {journal} {\bibinfo  {journal} {Physical Review A}\ }\textbf
  {\bibinfo {volume} {55}},\ \bibinfo {pages} {2989} (\bibinfo {year}
  {1997})}\BibitemShut {NoStop}%
\bibitem [{\citenamefont {Holthaus}(2000)}]{Holthaus2000}%
  \BibitemOpen
  \bibfield  {author} {\bibinfo {author} {\bibfnamefont {M.}~\bibnamefont
  {Holthaus}},\ }\bibfield  {title} {\emph {\bibinfo {title} {Bloch
  oscillations and {{Zener}} breakdown in an optical lattice},\ }}\href
  {\doibase10.1088/1464-4266/2/5/306} {\bibfield  {journal} {\bibinfo
  {journal} {Journal of Optics B: Quantum and Semiclassical Optics}\ }\textbf
  {\bibinfo {volume} {2}},\ \bibinfo {pages} {589} (\bibinfo {year}
  {2000})}\BibitemShut {NoStop}%
\bibitem [{\citenamefont {Sias}\ \emph {et~al.}(2007)\citenamefont {Sias},
  \citenamefont {Zenesini}, \citenamefont {Lignier}, \citenamefont {Wimberger},
  \citenamefont {Ciampini}, \citenamefont {Morsch},\ and\ \citenamefont
  {Arimondo}}]{Sias2007}%
  \BibitemOpen
  \bibfield  {author} {\bibinfo {author} {\bibfnamefont {C.}~\bibnamefont
  {Sias}}, \bibinfo {author} {\bibfnamefont {A.}~\bibnamefont {Zenesini}},
  \bibinfo {author} {\bibfnamefont {H.}~\bibnamefont {Lignier}}, \bibinfo
  {author} {\bibfnamefont {S.}~\bibnamefont {Wimberger}}, \bibinfo {author}
  {\bibfnamefont {D.}~\bibnamefont {Ciampini}}, \bibinfo {author}
  {\bibfnamefont {O.}~\bibnamefont {Morsch}}, \ and\ \bibinfo {author}
  {\bibfnamefont {E.}~\bibnamefont {Arimondo}},\ }\bibfield  {title} {\emph
  {\bibinfo {title} {Resonantly {{Enhanced Tunneling}} of {{Bose-Einstein
  Condensates}} in {{Periodic Potentials}}},\ }}\href
  {\doibase10.1103/PhysRevLett.98.120403} {\bibfield  {journal} {\bibinfo
  {journal} {Physical Review Letters}\ }\textbf {\bibinfo {volume} {98}},\
  \bibinfo {pages} {120403} (\bibinfo {year} {2007})}\BibitemShut {NoStop}%
\bibitem [{\citenamefont {Clad{\'e}}\ \emph {et~al.}(2017)\citenamefont
  {Clad{\'e}}, \citenamefont {Andia},\ and\ \citenamefont
  {{Guellati-Kh{\'e}lifa}}}]{Clade2017}%
  \BibitemOpen
  \bibfield  {author} {\bibinfo {author} {\bibfnamefont {P.}~\bibnamefont
  {Clad{\'e}}}, \bibinfo {author} {\bibfnamefont {M.}~\bibnamefont {Andia}}, \
  and\ \bibinfo {author} {\bibfnamefont {S.}~\bibnamefont
  {{Guellati-Kh{\'e}lifa}}},\ }\bibfield  {title} {\emph {\bibinfo {title}
  {Improving efficiency of {{Bloch}} oscillations in the tight-binding limit},\
  }}\href {\doibase10.1103/PhysRevA.95.063604} {\bibfield  {journal} {\bibinfo
  {journal} {Physical Review A}\ }\textbf {\bibinfo {volume} {95}},\ \bibinfo
  {pages} {063604} (\bibinfo {year} {2017})}\BibitemShut {NoStop}%
\bibitem [{Note1()}]{Note1}%
  \BibitemOpen
  \bibinfo {note} {For non-vanishing ramping time $\tau _{ramp}$, $T$ is
  determined by integrating the acceleration curve $a_L(t)$ and solving for the
  final velocity $2N\hbar k_L/m$.}\BibitemShut {Stop}%
\bibitem [{\citenamefont {Wannier}(1960)}]{Wannier1960}%
  \BibitemOpen
  \bibfield  {author} {\bibinfo {author} {\bibfnamefont {G.~H.}\ \bibnamefont
  {Wannier}},\ }\bibfield  {title} {\emph {\bibinfo {title} {Wave {{Functions}}
  and {{Effective Hamiltonian}} for {{Bloch Electrons}} in an {{Electric
  Field}}},\ }}\href {\doibase10.1103/PhysRev.117.432} {\bibfield  {journal}
  {\bibinfo  {journal} {Physical Review}\ }\textbf {\bibinfo {volume} {117}},\
  \bibinfo {pages} {432} (\bibinfo {year} {1960})}\BibitemShut {NoStop}%
\bibitem [{\citenamefont {Nenciu}(1991)}]{Nenciu1991}%
  \BibitemOpen
  \bibfield  {author} {\bibinfo {author} {\bibfnamefont {G.}~\bibnamefont
  {Nenciu}},\ }\bibfield  {title} {\emph {\bibinfo {title} {Dynamics of band
  electrons in electric and magnetic fields: Rigorous justification of the
  effective {{Hamiltonians}}},\ }}\href {\doibase10.1103/RevModPhys.63.91}
  {\bibfield  {journal} {\bibinfo  {journal} {Reviews of Modern Physics}\
  }\textbf {\bibinfo {volume} {63}},\ \bibinfo {pages} {91} (\bibinfo {year}
  {1991})}\BibitemShut {NoStop}%
\bibitem [{\citenamefont {Gl{\"u}ck}\ \emph {et~al.}(2002)\citenamefont
  {Gl{\"u}ck}, \citenamefont {R.~Kolovsky},\ and\ \citenamefont
  {Korsch}}]{Gluck2002}%
  \BibitemOpen
  \bibfield  {author} {\bibinfo {author} {\bibfnamefont {M.}~\bibnamefont
  {Gl{\"u}ck}}, \bibinfo {author} {\bibfnamefont {A.}~\bibnamefont
  {R.~Kolovsky}}, \ and\ \bibinfo {author} {\bibfnamefont {H.~J.}\ \bibnamefont
  {Korsch}},\ }\bibfield  {title} {\emph {\bibinfo {title}
  {Wannier\textendash{{Stark}} resonances in optical and semiconductor
  superlattices},\ }}\href {\doibase10.1016/S0370-1573(02)00142-4} {\bibfield
  {journal} {\bibinfo  {journal} {Physics Reports}\ }\textbf {\bibinfo {volume}
  {366}},\ \bibinfo {pages} {103} (\bibinfo {year} {2002})}\BibitemShut
  {NoStop}%
\bibitem [{\citenamefont {Zak}(1968)}]{Zak1968}%
  \BibitemOpen
  \bibfield  {author} {\bibinfo {author} {\bibfnamefont {J.}~\bibnamefont
  {Zak}},\ }\bibfield  {title} {\emph {\bibinfo {title} {Stark {{Ladder}} in
  {{Solids}}?}\ }}\href {\doibase10.1103/PhysRevLett.20.1477} {\bibfield
  {journal} {\bibinfo  {journal} {Physical Review Letters}\ }\textbf {\bibinfo
  {volume} {20}},\ \bibinfo {pages} {1477} (\bibinfo {year}
  {1968})}\BibitemShut {NoStop}%
\bibitem [{\citenamefont {Wannier}(1969)}]{Wannier1969}%
  \BibitemOpen
  \bibfield  {author} {\bibinfo {author} {\bibfnamefont {G.~H.}\ \bibnamefont
  {Wannier}},\ }\bibfield  {title} {\emph {\bibinfo {title} {Stark {{Ladder}}
  in {{Solids}}? {{A Reply}}},\ }}\href {\doibase10.1103/PhysRev.181.1364}
  {\bibfield  {journal} {\bibinfo  {journal} {Physical Review}\ }\textbf
  {\bibinfo {volume} {181}},\ \bibinfo {pages} {1364} (\bibinfo {year}
  {1969})}\BibitemShut {NoStop}%
\bibitem [{\citenamefont {Zak}(1969)}]{Zak1969}%
  \BibitemOpen
  \bibfield  {author} {\bibinfo {author} {\bibfnamefont {J.}~\bibnamefont
  {Zak}},\ }\bibfield  {title} {\emph {\bibinfo {title} {Stark {{Ladder}} in
  {{Solids}}? {{A Reply}} to a {{Reply}}},\ }}\href
  {\doibase10.1103/PhysRev.181.1366} {\bibfield  {journal} {\bibinfo  {journal}
  {Physical Review}\ }\textbf {\bibinfo {volume} {181}},\ \bibinfo {pages}
  {1366} (\bibinfo {year} {1969})}\BibitemShut {NoStop}%
\bibitem [{Note2()}]{Note2}%
  \BibitemOpen
  \bibinfo {note} {We discuss excited-band Bloch oscillations proposed by~\cite
  {McAlpine2020} in~\cite {SupplementFitzek2023}.}\BibitemShut {Stop}%
\bibitem [{\citenamefont {Mondrag{\'o}n}\ and\ \citenamefont
  {Hern{\'a}ndez}(1996)}]{Mondragon1996}%
  \BibitemOpen
  \bibfield  {author} {\bibinfo {author} {\bibfnamefont {A.}~\bibnamefont
  {Mondrag{\'o}n}}\ and\ \bibinfo {author} {\bibfnamefont {E.}~\bibnamefont
  {Hern{\'a}ndez}},\ }\bibfield  {title} {\emph {\bibinfo {title} {Berry phase
  of a resonant state},\ }}\href {\doibase10.1088/0305-4470/29/10/032}
  {\bibfield  {journal} {\bibinfo  {journal} {Journal of Physics A:
  Mathematical and General}\ }\textbf {\bibinfo {volume} {29}},\ \bibinfo
  {pages} {2567} (\bibinfo {year} {1996})}\BibitemShut {NoStop}%
\bibitem [{\citenamefont {Keck}\ \emph {et~al.}(2003)\citenamefont {Keck},
  \citenamefont {Korsch},\ and\ \citenamefont {Mossmann}}]{Keck2003}%
  \BibitemOpen
  \bibfield  {author} {\bibinfo {author} {\bibfnamefont {F.}~\bibnamefont
  {Keck}}, \bibinfo {author} {\bibfnamefont {H.~J.}\ \bibnamefont {Korsch}}, \
  and\ \bibinfo {author} {\bibfnamefont {S.}~\bibnamefont {Mossmann}},\
  }\bibfield  {title} {\emph {\bibinfo {title} {Unfolding a diabolic point: A
  generalized crossing scenario},\ }}\href {\doibase10.1088/0305-4470/36/8/310}
  {\bibfield  {journal} {\bibinfo  {journal} {Journal of Physics A:
  Mathematical and General}\ }\textbf {\bibinfo {volume} {36}},\ \bibinfo
  {pages} {2125} (\bibinfo {year} {2003})}\BibitemShut {NoStop}%
\bibitem [{\citenamefont {Ib{\'a}{\~n}ez}\ and\ \citenamefont
  {Muga}(2014)}]{Ibanez2014}%
  \BibitemOpen
  \bibfield  {author} {\bibinfo {author} {\bibfnamefont {S.}~\bibnamefont
  {Ib{\'a}{\~n}ez}}\ and\ \bibinfo {author} {\bibfnamefont {J.~G.}\
  \bibnamefont {Muga}},\ }\bibfield  {title} {\emph {\bibinfo {title}
  {Adiabaticity condition for non-{{Hermitian Hamiltonians}}},\ }}\href
  {\doibase10.1103/PhysRevA.89.033403} {\bibfield  {journal} {\bibinfo
  {journal} {Physical Review A}\ }\textbf {\bibinfo {volume} {89}},\ \bibinfo
  {pages} {033403} (\bibinfo {year} {2014})}\BibitemShut {NoStop}%
\bibitem [{\citenamefont {Feit}\ \emph {et~al.}(1982)\citenamefont {Feit},
  \citenamefont {Fleck},\ and\ \citenamefont {Steiger}}]{Feit1982}%
  \BibitemOpen
  \bibfield  {author} {\bibinfo {author} {\bibfnamefont {M.~D.}\ \bibnamefont
  {Feit}}, \bibinfo {author} {\bibfnamefont {J.~A.}\ \bibnamefont {Fleck}}, \
  and\ \bibinfo {author} {\bibfnamefont {A.}~\bibnamefont {Steiger}},\
  }\bibfield  {title} {\emph {\bibinfo {title} {Solution of the
  {{Schr\"odinger}} equation by a spectral method},\ }}\href
  {\doibase10.1016/0021-9991(82)90091-2} {\bibfield  {journal} {\bibinfo
  {journal} {Journal of Computational Physics}\ }\textbf {\bibinfo {volume}
  {47}},\ \bibinfo {pages} {412} (\bibinfo {year} {1982})}\BibitemShut
  {NoStop}%
\bibitem [{\citenamefont {Fitzek}\ \emph {et~al.}(2020)\citenamefont {Fitzek},
  \citenamefont {Siem{\ss}}, \citenamefont {Seckmeyer}, \citenamefont {Ahlers},
  \citenamefont {Rasel}, \citenamefont {Hammerer},\ and\ \citenamefont
  {Gaaloul}}]{Fitzek2020}%
  \BibitemOpen
  \bibfield  {author} {\bibinfo {author} {\bibfnamefont {F.}~\bibnamefont
  {Fitzek}}, \bibinfo {author} {\bibfnamefont {J.-N.}\ \bibnamefont
  {Siem{\ss}}}, \bibinfo {author} {\bibfnamefont {S.}~\bibnamefont
  {Seckmeyer}}, \bibinfo {author} {\bibfnamefont {H.}~\bibnamefont {Ahlers}},
  \bibinfo {author} {\bibfnamefont {E.~M.}\ \bibnamefont {Rasel}}, \bibinfo
  {author} {\bibfnamefont {K.}~\bibnamefont {Hammerer}}, \ and\ \bibinfo
  {author} {\bibfnamefont {N.}~\bibnamefont {Gaaloul}},\ }\bibfield  {title}
  {\emph {\bibinfo {title} {Universal atom interferometer simulation of elastic
  scattering processes},\ }}\href {\doibase10.1038/s41598-020-78859-1}
  {\bibfield  {journal} {\bibinfo  {journal} {Scientific Reports}\ }\textbf
  {\bibinfo {volume} {10}},\ \bibinfo {pages} {22120} (\bibinfo {year}
  {2020})}\BibitemShut {NoStop}%
\bibitem [{\citenamefont {Muga}\ \emph {et~al.}(2004)\citenamefont {Muga},
  \citenamefont {Palao}, \citenamefont {Navarro},\ and\ \citenamefont
  {Egusquiza}}]{Muga2004}%
  \BibitemOpen
  \bibfield  {author} {\bibinfo {author} {\bibfnamefont {J.~G.}\ \bibnamefont
  {Muga}}, \bibinfo {author} {\bibfnamefont {J.~P.}\ \bibnamefont {Palao}},
  \bibinfo {author} {\bibfnamefont {B.}~\bibnamefont {Navarro}}, \ and\
  \bibinfo {author} {\bibfnamefont {I.~L.}\ \bibnamefont {Egusquiza}},\
  }\bibfield  {title} {\emph {\bibinfo {title} {Complex absorbing potentials},\
  }}\href {\doibase10.1016/j.physrep.2004.03.002} {\bibfield  {journal}
  {\bibinfo  {journal} {Physics Reports}\ }\textbf {\bibinfo {volume} {395}},\
  \bibinfo {pages} {357} (\bibinfo {year} {2004})}\BibitemShut {NoStop}%
\bibitem [{Note3()}]{Note3}%
  \BibitemOpen
  \bibinfo {note} {We discuss the lattice-shift method proposed by~\cite
  {Clade2017} in~\cite {SupplementFitzek2023}.}\BibitemShut {Stop}%
\bibitem [{\citenamefont {Kim}\ \emph {et~al.}(2020)\citenamefont {Kim},
  \citenamefont {Notermans}, \citenamefont {Overstreet}, \citenamefont {Curti},
  \citenamefont {Asenbaum},\ and\ \citenamefont {Kasevich}}]{Kim2020}%
  \BibitemOpen
  \bibfield  {author} {\bibinfo {author} {\bibfnamefont {M.}~\bibnamefont
  {Kim}}, \bibinfo {author} {\bibfnamefont {R.}~\bibnamefont {Notermans}},
  \bibinfo {author} {\bibfnamefont {C.}~\bibnamefont {Overstreet}}, \bibinfo
  {author} {\bibfnamefont {J.}~\bibnamefont {Curti}}, \bibinfo {author}
  {\bibfnamefont {P.}~\bibnamefont {Asenbaum}}, \ and\ \bibinfo {author}
  {\bibfnamefont {M.~A.}\ \bibnamefont {Kasevich}},\ }\bibfield  {title} {\emph
  {\bibinfo {title} {40 {{W}}, 780 nm laser system with compensated dual beam
  splitters for atom interferometry},\ }}\href {\doibase10.1364/OL.404430}
  {\bibfield  {journal} {\bibinfo  {journal} {Optics Letters}\ }\textbf
  {\bibinfo {volume} {45}},\ \bibinfo {pages} {6555} (\bibinfo {year}
  {2020})}\BibitemShut {NoStop}%
\bibitem [{\citenamefont {Panda}\ \emph
  {et~al.}(2022{\natexlab{b}})\citenamefont {Panda}, \citenamefont {Tao},
  \citenamefont {Egelhoff}, \citenamefont {Ceja}, \citenamefont {Xu},\ and\
  \citenamefont {M{\"u}ller}}]{Panda2022a}%
  \BibitemOpen
  \bibfield  {author} {\bibinfo {author} {\bibfnamefont {C.~D.}\ \bibnamefont
  {Panda}}, \bibinfo {author} {\bibfnamefont {M.}~\bibnamefont {Tao}}, \bibinfo
  {author} {\bibfnamefont {J.}~\bibnamefont {Egelhoff}}, \bibinfo {author}
  {\bibfnamefont {M.}~\bibnamefont {Ceja}}, \bibinfo {author} {\bibfnamefont
  {V.}~\bibnamefont {Xu}}, \ and\ \bibinfo {author} {\bibfnamefont
  {H.}~\bibnamefont {M{\"u}ller}},\ }\href@noop {} {\bibinfo {title} {Quantum
  metrology by one-minute interrogation of a coherent atomic spatial
  superposition},\ } (\bibinfo {year} {2022}{\natexlab{b}}),\ \bibinfo {note}
  {comment: 23 pages, 9 figures},\ \Eprint {http://arxiv.org/abs/2210.07289}
  {arxiv:2210.07289 [gr-qc, physics:physics, physics:quant-ph]} \BibitemShut
  {NoStop}%
\bibitem [{\citenamefont {Morel}(2019)}]{Morel2019}%
  \BibitemOpen
  \bibfield  {author} {\bibinfo {author} {\bibfnamefont {L.}~\bibnamefont
  {Morel}},\ }\emph {\bibinfo {title} {High Sensitivity Matter-Wave
  Interferometry : Towards a Determination of the Fine Structure Constant below
  10-10}},\ \href@noop {} {Ph.D. thesis},\ \bibinfo  {school} {Sorbonne
  Universit\'e} (\bibinfo {year} {2019})\BibitemShut {NoStop}%
\bibitem [{\citenamefont {Canuel}\ \emph
  {et~al.}(2020{\natexlab{b}})\citenamefont {Canuel} \emph
  {et~al.}}]{Canuel2020a}%
  \BibitemOpen
  \bibfield  {author} {\bibinfo {author} {\bibfnamefont {B.}~\bibnamefont
  {Canuel}} \emph {et~al.},\ }\href {\doibase10.48550/arXiv.2007.04014}
  {\bibinfo {title} {Technologies for the {{ELGAR}} large scale atom
  interferometer array},\ } (\bibinfo {year} {2020}{\natexlab{b}}),\ \Eprint
  {http://arxiv.org/abs/2007.04014} {arxiv:2007.04014 [physics]} \BibitemShut
  {NoStop}%
\bibitem [{\citenamefont {Martin}\ \emph {et~al.}(1988)\citenamefont {Martin},
  \citenamefont {Oldaker}, \citenamefont {Miklich},\ and\ \citenamefont
  {Pritchard}}]{Martin1988}%
  \BibitemOpen
  \bibfield  {author} {\bibinfo {author} {\bibfnamefont {P.~J.}\ \bibnamefont
  {Martin}}, \bibinfo {author} {\bibfnamefont {B.~G.}\ \bibnamefont {Oldaker}},
  \bibinfo {author} {\bibfnamefont {A.~H.}\ \bibnamefont {Miklich}}, \ and\
  \bibinfo {author} {\bibfnamefont {D.~E.}\ \bibnamefont {Pritchard}},\
  }\bibfield  {title} {\emph {\bibinfo {title} {Bragg scattering of atoms from
  a standing light wave},\ }}\href {\doibase10.1103/PhysRevLett.60.515}
  {\bibfield  {journal} {\bibinfo  {journal} {Physical Review Letters}\
  }\textbf {\bibinfo {volume} {60}},\ \bibinfo {pages} {515} (\bibinfo {year}
  {1988})}\BibitemShut {NoStop}%
\bibitem [{\citenamefont {Giltner}\ \emph {et~al.}(1995)\citenamefont
  {Giltner}, \citenamefont {McGowan},\ and\ \citenamefont
  {Lee}}]{Giltner1995PRA}%
  \BibitemOpen
  \bibfield  {author} {\bibinfo {author} {\bibfnamefont {D.~M.}\ \bibnamefont
  {Giltner}}, \bibinfo {author} {\bibfnamefont {R.~W.}\ \bibnamefont
  {McGowan}}, \ and\ \bibinfo {author} {\bibfnamefont {S.~A.}\ \bibnamefont
  {Lee}},\ }\bibfield  {title} {\emph {\bibinfo {title} {Theoretical and
  experimental study of the {{Bragg}} scattering of atoms from a standing light
  wave},\ }}\href {\doibase10.1103/PhysRevA.52.3966} {\bibfield  {journal}
  {\bibinfo  {journal} {Physical Review A}\ }\textbf {\bibinfo {volume} {52}},\
  \bibinfo {pages} {3966} (\bibinfo {year} {1995})}\BibitemShut {NoStop}%
\bibitem [{\citenamefont {Cheinet}\ \emph {et~al.}(2008)\citenamefont
  {Cheinet}, \citenamefont {Canuel}, \citenamefont {Pereira Dos~Santos},
  \citenamefont {Gauguet}, \citenamefont {{Yver-Leduc}},\ and\ \citenamefont
  {Landragin}}]{Cheinet2008}%
  \BibitemOpen
  \bibfield  {author} {\bibinfo {author} {\bibfnamefont {P.}~\bibnamefont
  {Cheinet}}, \bibinfo {author} {\bibfnamefont {B.}~\bibnamefont {Canuel}},
  \bibinfo {author} {\bibfnamefont {F.}~\bibnamefont {Pereira Dos~Santos}},
  \bibinfo {author} {\bibfnamefont {A.}~\bibnamefont {Gauguet}}, \bibinfo
  {author} {\bibfnamefont {F.}~\bibnamefont {{Yver-Leduc}}}, \ and\ \bibinfo
  {author} {\bibfnamefont {A.}~\bibnamefont {Landragin}},\ }\bibfield  {title}
  {\emph {\bibinfo {title} {Measurement of the {{Sensitivity Function}} in a
  {{Time-Domain Atomic Interferometer}}},\ }}\href
  {\doibase10.1109/TIM.2007.915148} {\bibfield  {journal} {\bibinfo  {journal}
  {IEEE Transactions on Instrumentation and Measurement}\ }\textbf {\bibinfo
  {volume} {57}},\ \bibinfo {pages} {1141} (\bibinfo {year}
  {2008})}\BibitemShut {NoStop}%
\bibitem [{\citenamefont {Pichler}\ \emph {et~al.}(2010)\citenamefont
  {Pichler}, \citenamefont {Daley},\ and\ \citenamefont
  {Zoller}}]{Pichler2010}%
  \BibitemOpen
  \bibfield  {author} {\bibinfo {author} {\bibfnamefont {H.}~\bibnamefont
  {Pichler}}, \bibinfo {author} {\bibfnamefont {A.~J.}\ \bibnamefont {Daley}},
  \ and\ \bibinfo {author} {\bibfnamefont {P.}~\bibnamefont {Zoller}},\
  }\bibfield  {title} {\emph {\bibinfo {title} {Nonequilibrium dynamics of
  bosonic atoms in optical lattices: {{Decoherence}} of many-body states due to
  spontaneous emission},\ }}\href {\doibase10.1103/PhysRevA.82.063605}
  {\bibfield  {journal} {\bibinfo  {journal} {Physical Review A}\ }\textbf
  {\bibinfo {volume} {82}},\ \bibinfo {pages} {063605} (\bibinfo {year}
  {2010})}\BibitemShut {NoStop}%
\bibitem [{\citenamefont {Gl{\"u}ck}\ \emph {et~al.}(2001)\citenamefont
  {Gl{\"u}ck}, \citenamefont {Keck}, \citenamefont {Kolovsky},\ and\
  \citenamefont {Korsch}}]{Gluck2001}%
  \BibitemOpen
  \bibfield  {author} {\bibinfo {author} {\bibfnamefont {M.}~\bibnamefont
  {Gl{\"u}ck}}, \bibinfo {author} {\bibfnamefont {F.}~\bibnamefont {Keck}},
  \bibinfo {author} {\bibfnamefont {A.~R.}\ \bibnamefont {Kolovsky}}, \ and\
  \bibinfo {author} {\bibfnamefont {H.~J.}\ \bibnamefont {Korsch}},\ }\bibfield
   {title} {\emph {\bibinfo {title} {Wannier-{{Stark States}} of a {{Quantum
  Particle}} in {{2D Lattices}}},\ }}\href
  {\doibase10.1103/PhysRevLett.86.3116} {\bibfield  {journal} {\bibinfo
  {journal} {Physical Review Letters}\ }\textbf {\bibinfo {volume} {86}},\
  \bibinfo {pages} {3116} (\bibinfo {year} {2001})}\BibitemShut {NoStop}%
\bibitem [{Note4()}]{Note4}%
  \BibitemOpen
  \bibinfo {note} {T. Rahman et al., in preperation}\BibitemShut {NoStop}%
\bibitem [{\citenamefont {McAlpine}\ \emph {et~al.}(2020)\citenamefont
  {McAlpine}, \citenamefont {Gochnauer},\ and\ \citenamefont
  {Gupta}}]{McAlpine2020}%
  \BibitemOpen
  \bibfield  {author} {\bibinfo {author} {\bibfnamefont {K.~E.}\ \bibnamefont
  {McAlpine}}, \bibinfo {author} {\bibfnamefont {D.}~\bibnamefont {Gochnauer}},
  \ and\ \bibinfo {author} {\bibfnamefont {S.}~\bibnamefont {Gupta}},\
  }\bibfield  {title} {\emph {\bibinfo {title} {Excited-band {{Bloch}}
  oscillations for precision atom interferometry},\ }}\href
  {\doibase10.1103/PhysRevA.101.023614} {\bibfield  {journal} {\bibinfo
  {journal} {Physical Review A}\ }\textbf {\bibinfo {volume} {101}},\ \bibinfo
  {pages} {023614} (\bibinfo {year} {2020})}\BibitemShut {NoStop}%
\end{thebibliography}%


\begin{thebibliography}{14}%
\makeatletter
\providecommand \@ifxundefined [1]{%
 \@ifx{#1\undefined}
}%
\providecommand \@ifnum [1]{%
 \ifnum #1\expandafter \@firstoftwo
 \else \expandafter \@secondoftwo
 \fi
}%
\providecommand \@ifx [1]{%
 \ifx #1\expandafter \@firstoftwo
 \else \expandafter \@secondoftwo
 \fi
}%
\providecommand \natexlab [1]{#1}%
\providecommand \enquote  [1]{``#1''}%
\providecommand \bibnamefont  [1]{#1}%
\providecommand \bibfnamefont [1]{#1}%
\providecommand \citenamefont [1]{#1}%
\providecommand \href@noop [0]{\@secondoftwo}%
\providecommand \href [0]{\begingroup \@sanitize@url \@href}%
\providecommand \@href[1]{\@@startlink{#1}\@@href}%
\providecommand \@@href[1]{\endgroup#1\@@endlink}%
\providecommand \@sanitize@url [0]{\catcode `\\12\catcode `\$12\catcode
  `\&12\catcode `\#12\catcode `\^12\catcode `\_12\catcode `\%12\relax}%
\providecommand \@@startlink[1]{}%
\providecommand \@@endlink[0]{}%
\providecommand \url  [0]{\begingroup\@sanitize@url \@url }%
\providecommand \@url [1]{\endgroup\@href {#1}{\urlprefix }}%
\providecommand \urlprefix  [0]{URL }%
\providecommand \Eprint [0]{\href }%
\providecommand \doibase [0]{http://dx.doi.org/}%
\providecommand \selectlanguage [0]{\@gobble}%
\providecommand \bibinfo  [0]{\@secondoftwo}%
\providecommand \bibfield  [0]{\@secondoftwo}%
\providecommand \translation [1]{[#1]}%
\providecommand \BibitemOpen [0]{}%
\providecommand \bibitemStop [0]{}%
\providecommand \bibitemNoStop [0]{.\EOS\space}%
\providecommand \EOS [0]{\spacefactor3000\relax}%
\providecommand \BibitemShut  [1]{\csname bibitem#1\endcsname}%
\let\auto@bib@innerbib\@empty
%</preamble>
\bibitem [{\citenamefont {Bloch}(1929)}]{Bloch1929}%
  \BibitemOpen
  \bibfield  {author} {\bibinfo {author} {\bibfnamefont {F.}~\bibnamefont
  {Bloch}},\ }\bibfield  {title} {\emph {\bibinfo {title} {{\"Uber die
  Quantenmechanik der Elektronen in Kristallgittern}},\ }}\href
  {\doibase10.1007/BF01339455} {\bibfield  {journal} {\bibinfo  {journal}
  {Zeitschrift f\"ur Physik}\ }\textbf {\bibinfo {volume} {52}},\ \bibinfo
  {pages} {555} (\bibinfo {year} {1929})}\BibitemShut {NoStop}%
\bibitem [{\citenamefont {Zener}\ and\ \citenamefont
  {Fowler}(1934)}]{Zener1934}%
  \BibitemOpen
  \bibfield  {author} {\bibinfo {author} {\bibfnamefont {C.}~\bibnamefont
  {Zener}}\ and\ \bibinfo {author} {\bibfnamefont {R.~H.}\ \bibnamefont
  {Fowler}},\ }\bibfield  {title} {\emph {\bibinfo {title} {A theory of the
  electrical breakdown of solid dielectrics},\ }}\href
  {\doibase10.1098/rspa.1934.0116} {\bibfield  {journal} {\bibinfo  {journal}
  {Proceedings of the Royal Society of London. Series A, Containing Papers of a
  Mathematical and Physical Character}\ }\textbf {\bibinfo {volume} {145}},\
  \bibinfo {pages} {523} (\bibinfo {year} {1934})}\BibitemShut {NoStop}%
\bibitem [{\citenamefont {Gl{\"u}ck}\ \emph {et~al.}(2002)\citenamefont
  {Gl{\"u}ck}, \citenamefont {R.~Kolovsky},\ and\ \citenamefont
  {Korsch}}]{Gluck2002}%
  \BibitemOpen
  \bibfield  {author} {\bibinfo {author} {\bibfnamefont {M.}~\bibnamefont
  {Gl{\"u}ck}}, \bibinfo {author} {\bibfnamefont {A.}~\bibnamefont
  {R.~Kolovsky}}, \ and\ \bibinfo {author} {\bibfnamefont {H.~J.}\ \bibnamefont
  {Korsch}},\ }\bibfield  {title} {\emph {\bibinfo {title}
  {Wannier\textendash{{Stark}} resonances in optical and semiconductor
  superlattices},\ }}\href {\doibase10.1016/S0370-1573(02)00142-4} {\bibfield
  {journal} {\bibinfo  {journal} {Physics Reports}\ }\textbf {\bibinfo {volume}
  {366}},\ \bibinfo {pages} {103} (\bibinfo {year} {2002})}\BibitemShut
  {NoStop}%
\bibitem [{\citenamefont {Gl{\"u}ck}\ \emph {et~al.}(1998)\citenamefont
  {Gl{\"u}ck}, \citenamefont {Kolovsky}, \citenamefont {Korsch},\ and\
  \citenamefont {Moiseyev}}]{Gluck1998}%
  \BibitemOpen
  \bibfield  {author} {\bibinfo {author} {\bibfnamefont {M.}~\bibnamefont
  {Gl{\"u}ck}}, \bibinfo {author} {\bibfnamefont {A.}~\bibnamefont {Kolovsky}},
  \bibinfo {author} {\bibfnamefont {H.}~\bibnamefont {Korsch}}, \ and\ \bibinfo
  {author} {\bibfnamefont {N.}~\bibnamefont {Moiseyev}},\ }\bibfield  {title}
  {\emph {\bibinfo {title} {Calculation of {{Wannier-Bloch}} and
  {{Wannier-Stark}} states},\ }}\href {\doibase10.1007/s100530050205}
  {\bibfield  {journal} {\bibinfo  {journal} {The European Physical Journal D -
  Atomic, Molecular, Optical and Plasma Physics}\ }\textbf {\bibinfo {volume}
  {4}},\ \bibinfo {pages} {239} (\bibinfo {year} {1998})}\BibitemShut {NoStop}%
\bibitem [{\citenamefont {Virtanen}\ \emph {et~al.}(2020)\citenamefont
  {Virtanen} \emph {et~al.}}]{Virtanen2020}%
  \BibitemOpen
  \bibfield  {author} {\bibinfo {author} {\bibfnamefont {P.}~\bibnamefont
  {Virtanen}} \emph {et~al.},\ }\bibfield  {title} {\emph {\bibinfo {title}
  {{{SciPy}} 1.0: Fundamental algorithms for scientific computing in
  {{Python}}},\ }}\href {\doibase10.1038/s41592-019-0686-2} {\bibfield
  {journal} {\bibinfo  {journal} {Nature Methods}\ }\textbf {\bibinfo {volume}
  {17}},\ \bibinfo {pages} {261} (\bibinfo {year} {2020})}\BibitemShut
  {NoStop}%
\bibitem [{\citenamefont {Mondrag{\'o}n}\ and\ \citenamefont
  {Hern{\'a}ndez}(1996)}]{Mondragon1996}%
  \BibitemOpen
  \bibfield  {author} {\bibinfo {author} {\bibfnamefont {A.}~\bibnamefont
  {Mondrag{\'o}n}}\ and\ \bibinfo {author} {\bibfnamefont {E.}~\bibnamefont
  {Hern{\'a}ndez}},\ }\bibfield  {title} {\emph {\bibinfo {title} {Berry phase
  of a resonant state},\ }}\href {\doibase10.1088/0305-4470/29/10/032}
  {\bibfield  {journal} {\bibinfo  {journal} {Journal of Physics A:
  Mathematical and General}\ }\textbf {\bibinfo {volume} {29}},\ \bibinfo
  {pages} {2567} (\bibinfo {year} {1996})}\BibitemShut {NoStop}%
\bibitem [{\citenamefont {Keck}\ \emph {et~al.}(2003)\citenamefont {Keck},
  \citenamefont {Korsch},\ and\ \citenamefont {Mossmann}}]{Keck2003}%
  \BibitemOpen
  \bibfield  {author} {\bibinfo {author} {\bibfnamefont {F.}~\bibnamefont
  {Keck}}, \bibinfo {author} {\bibfnamefont {H.~J.}\ \bibnamefont {Korsch}}, \
  and\ \bibinfo {author} {\bibfnamefont {S.}~\bibnamefont {Mossmann}},\
  }\bibfield  {title} {\emph {\bibinfo {title} {Unfolding a diabolic point: A
  generalized crossing scenario},\ }}\href {\doibase10.1088/0305-4470/36/8/310}
  {\bibfield  {journal} {\bibinfo  {journal} {Journal of Physics A:
  Mathematical and General}\ }\textbf {\bibinfo {volume} {36}},\ \bibinfo
  {pages} {2125} (\bibinfo {year} {2003})}\BibitemShut {NoStop}%
\bibitem [{\citenamefont {Ib{\'a}{\~n}ez}\ and\ \citenamefont
  {Muga}(2014)}]{Ibanez2014}%
  \BibitemOpen
  \bibfield  {author} {\bibinfo {author} {\bibfnamefont {S.}~\bibnamefont
  {Ib{\'a}{\~n}ez}}\ and\ \bibinfo {author} {\bibfnamefont {J.~G.}\
  \bibnamefont {Muga}},\ }\bibfield  {title} {\emph {\bibinfo {title}
  {Adiabaticity condition for non-{{Hermitian Hamiltonians}}},\ }}\href
  {\doibase10.1103/PhysRevA.89.033403} {\bibfield  {journal} {\bibinfo
  {journal} {Physical Review A}\ }\textbf {\bibinfo {volume} {89}},\ \bibinfo
  {pages} {033403} (\bibinfo {year} {2014})}\BibitemShut {NoStop}%
\bibitem [{\citenamefont {Grimm}\ \emph {et~al.}(2000)\citenamefont {Grimm},
  \citenamefont {Weidem{\"u}ller},\ and\ \citenamefont
  {Ovchinnikov}}]{Grimm2000}%
  \BibitemOpen
  \bibfield  {author} {\bibinfo {author} {\bibfnamefont {R.}~\bibnamefont
  {Grimm}}, \bibinfo {author} {\bibfnamefont {M.}~\bibnamefont
  {Weidem{\"u}ller}}, \ and\ \bibinfo {author} {\bibfnamefont {Y.~B.}\
  \bibnamefont {Ovchinnikov}},\ }in\ \href
  {\doibase10.1016/S1049-250X(08)60186-X} {\emph {\bibinfo {booktitle}
  {Advances {{In Atomic}}, {{Molecular}}, and {{Optical Physics}}}}},\
  Vol.~\bibinfo {volume} {42},\ \bibinfo {editor} {edited by\ \bibinfo {editor}
  {\bibfnamefont {B.}~\bibnamefont {Bederson}}\ and\ \bibinfo {editor}
  {\bibfnamefont {H.}~\bibnamefont {Walther}}}\ (\bibinfo  {publisher}
  {{Academic Press}},\ \bibinfo {year} {2000})\ pp.\ \bibinfo {pages}
  {95--170}\BibitemShut {NoStop}%
\bibitem [{\citenamefont {Gebbe}\ \emph {et~al.}(2021)\citenamefont {Gebbe}
  \emph {et~al.}}]{Gebbe2021}%
  \BibitemOpen
  \bibfield  {author} {\bibinfo {author} {\bibfnamefont {M.}~\bibnamefont
  {Gebbe}} \emph {et~al.},\ }\bibfield  {title} {\emph {\bibinfo {title}
  {Twin-lattice atom interferometry},\ }}\href
  {\doibase10.1038/s41467-021-22823-8} {\bibfield  {journal} {\bibinfo
  {journal} {Nature Communications}\ }\textbf {\bibinfo {volume} {12}},\
  \bibinfo {pages} {2544} (\bibinfo {year} {2021})}\BibitemShut {NoStop}%
\bibitem [{\citenamefont {Morel}\ \emph {et~al.}(2020)\citenamefont {Morel},
  \citenamefont {Yao}, \citenamefont {Clad{\'e}},\ and\ \citenamefont
  {{Guellati-Kh{\'e}lifa}}}]{Morel2020}%
  \BibitemOpen
  \bibfield  {author} {\bibinfo {author} {\bibfnamefont {L.}~\bibnamefont
  {Morel}}, \bibinfo {author} {\bibfnamefont {Z.}~\bibnamefont {Yao}}, \bibinfo
  {author} {\bibfnamefont {P.}~\bibnamefont {Clad{\'e}}}, \ and\ \bibinfo
  {author} {\bibfnamefont {S.}~\bibnamefont {{Guellati-Kh{\'e}lifa}}},\
  }\bibfield  {title} {\emph {\bibinfo {title} {Determination of the
  fine-structure constant with an accuracy of 81 parts per trillion},\ }}\href
  {\doibase10.1038/s41586-020-2964-7} {\bibfield  {journal} {\bibinfo
  {journal} {Nature}\ }\textbf {\bibinfo {volume} {588}},\ \bibinfo {pages}
  {61} (\bibinfo {year} {2020})}\BibitemShut {NoStop}%
\bibitem [{\citenamefont {Kim}\ \emph {et~al.}(2020)\citenamefont {Kim},
  \citenamefont {Notermans}, \citenamefont {Overstreet}, \citenamefont {Curti},
  \citenamefont {Asenbaum},\ and\ \citenamefont {Kasevich}}]{Kim2020}%
  \BibitemOpen
  \bibfield  {author} {\bibinfo {author} {\bibfnamefont {M.}~\bibnamefont
  {Kim}}, \bibinfo {author} {\bibfnamefont {R.}~\bibnamefont {Notermans}},
  \bibinfo {author} {\bibfnamefont {C.}~\bibnamefont {Overstreet}}, \bibinfo
  {author} {\bibfnamefont {J.}~\bibnamefont {Curti}}, \bibinfo {author}
  {\bibfnamefont {P.}~\bibnamefont {Asenbaum}}, \ and\ \bibinfo {author}
  {\bibfnamefont {M.~A.}\ \bibnamefont {Kasevich}},\ }\bibfield  {title} {\emph
  {\bibinfo {title} {40 {{W}}, 780 nm laser system with compensated dual beam
  splitters for atom interferometry},\ }}\href {\doibase10.1364/OL.404430}
  {\bibfield  {journal} {\bibinfo  {journal} {Optics Letters}\ }\textbf
  {\bibinfo {volume} {45}},\ \bibinfo {pages} {6555} (\bibinfo {year}
  {2020})}\BibitemShut {NoStop}%
\bibitem [{\citenamefont {Clad{\'e}}\ \emph {et~al.}(2017)\citenamefont
  {Clad{\'e}}, \citenamefont {Andia},\ and\ \citenamefont
  {{Guellati-Kh{\'e}lifa}}}]{Clade2017}%
  \BibitemOpen
  \bibfield  {author} {\bibinfo {author} {\bibfnamefont {P.}~\bibnamefont
  {Clad{\'e}}}, \bibinfo {author} {\bibfnamefont {M.}~\bibnamefont {Andia}}, \
  and\ \bibinfo {author} {\bibfnamefont {S.}~\bibnamefont
  {{Guellati-Kh{\'e}lifa}}},\ }\bibfield  {title} {\emph {\bibinfo {title}
  {Improving efficiency of {{Bloch}} oscillations in the tight-binding limit},\
  }}\href {\doibase10.1103/PhysRevA.95.063604} {\bibfield  {journal} {\bibinfo
  {journal} {Physical Review A}\ }\textbf {\bibinfo {volume} {95}},\ \bibinfo
  {pages} {063604} (\bibinfo {year} {2017})}\BibitemShut {NoStop}%
\bibitem [{\citenamefont {McAlpine}\ \emph {et~al.}(2020)\citenamefont
  {McAlpine}, \citenamefont {Gochnauer},\ and\ \citenamefont
  {Gupta}}]{McAlpine2020}%
  \BibitemOpen
  \bibfield  {author} {\bibinfo {author} {\bibfnamefont {K.~E.}\ \bibnamefont
  {McAlpine}}, \bibinfo {author} {\bibfnamefont {D.}~\bibnamefont {Gochnauer}},
  \ and\ \bibinfo {author} {\bibfnamefont {S.}~\bibnamefont {Gupta}},\
  }\bibfield  {title} {\emph {\bibinfo {title} {Excited-band {{Bloch}}
  oscillations for precision atom interferometry},\ }}\href
  {\doibase10.1103/PhysRevA.101.023614} {\bibfield  {journal} {\bibinfo
  {journal} {Physical Review A}\ }\textbf {\bibinfo {volume} {101}},\ \bibinfo
  {pages} {023614} (\bibinfo {year} {2020})}\BibitemShut {NoStop}%
\end{thebibliography}%

%%%%%%%%%%%%%%%%%%%%%%%%%%%%%%%%%%%%%%%%%%%%%%%%%%%%%%%%%%%%%%%%
%%%     End Document
%%%%%%%%%%%%%%%%%%%%%%%%%%%%%%%%%%%%%%%%%%%%%%%%%%%%%%%%%%%%%%%%

\end{document}

\title{Supplemental material: Accurate and efficient Bloch-oscillation-enhanced atom interferometry}

\def\affitp  {\affiliation{Leibniz Universit\"at Hannover, Institut f\"ur Theoretische Physik, Appelstr. 2, D-30167 Hannover, Germany }}
\def\affiqo  {\affiliation{Leibniz Universit\"at Hannover, Institut f\"ur Quantenoptik, Welfengarten 1, D-30167 Hannover, Germany }}
\def\affitp  {\affiliation{Leibniz Universit\"at Hannover, Institut f\"ur Theoretische Physik, Appelstr. 2, D-30167 Hannover, Germany }}
\def\affiqo  {\affiliation{Leibniz Universit\"at Hannover, Institut f\"ur Quantenoptik, Welfengarten 1, D-30167 Hannover, Germany }}
\author{F.~Fitzek}
\email[email: ]{fitzek@iqo.uni-hannover.de}
\affitp
\affiqo
\author{J.-N.~Kirsten-Siem\ss}
\affitp
\affiqo
\author{E.~M.~Rasel}
\author{N.~Gaaloul}
\affiqo
\author{K.~Hammerer}
\email[email: ]{klemens.hammerer@itp.uni-hannover.de}
\affitp
\date{\today}

\maketitle

\tableofcontents

\section{Hamiltonian and control parameters}\label{App:Hamiltonian and control parameters}
In this section, we present an overview of the unitary frames relevant to the discussion of LMT Bloch pulses and the Wannier-Stark model, as developed in the main article.
\subsection{Laboratory frame}
Starting from the laboratory frame, the Hamiltonian has the following form 
\begin{align} \label{eq:HamiltonianLab} 
        H_{\mathrm{lab\; frame}}(t)&=\frac{\hat{p}^2}{2m}+V_0(t) \cos^2(k_L (\hat{x}-x_L(t))),\\
        x_L(t)&=\int_0^t\mathrm{d}\tau\int_0^{\tau}\mathrm{d}\tau' a_L(\tau'),
\end{align}
where the control parameters $V_0(t)$ and $a_L(t)$ are piece-wise defined by

\begin{align}
    V_0(t) = 
     \begin{cases}
       \dfrac{V_0}{1+e^{-(t-\tau_{load}/2)/h\,\tau_{load}}} & 0<t<\tau_{load}\\
       V_0 & \tau_{load}<t<\tau_{load}+T \\
       \dfrac{V_0}{1+e^{(t-(T+3\tau_{load}/2))/h\,\tau_{load}}} & t>\tau_{load}+T
     \end{cases}
\end{align}
    
\begin{align}
     a_L(t) = 
     \begin{cases}
       0 & 0<t<\tau_{load}\\
       \dfrac{a_L}{1+e^{-(t-(\tau_{load}+\tau_{ramp}/2))/h\,\tau_{ramp}}} &\tau_{load}<t<\tau_{load}+\tau_{ramp} \\
       a_L &\tau_{load}+\tau_{ramp}<t<\tau_{load}+T-\tau_{ramp} \\
       \dfrac{a_L}{1+e^{(t-(\tau_{load}+T-\tau_{ramp}/2)/h\,\tau_{ramp}}} & \tau_{load}+T-\tau_{ramp}<t<\tau_{load}+T \\
       0 & t>\tau_{load}+T 
     \end{cases}
\end{align}    
as depicted in Fig.~1 (d) of the main article. By choosing a suitable value for the parameter $h$ we establish smooth connections of the sigmoidal rise and fall with the regions of constant acceleration and lattice depth. For our purposes, it is always set to $h=0.02$. 
\subsection{Lattice frame}
We transform to the comoving lattice frame with velocity $v_L(t)=p_L(t)/m$ and position $x_L(t)$ using the time-dependent unitary transformation
\begin{align} \label{eq:unitaryLattice} 
        U(t)&=\exp(i\hat{p}x_L(t)/\hbar-i\hat{x}p_L(t)/\hbar+i\Phi_U(t))
\end{align}
with an additional phase factor $\Phi_U(t)$, satisfying $\dot{\Phi}_U(t)=p_L^2(t)/2m$ to absorb a shift in the kinetic energy. Applying the following identities 
\begin{align} \label{eq:Hamiltonian} 
        e^{\hat{X}}\hat{Y}e^{-\hat{X}}&=\sum_{n=0}^{\infty}\frac{[\hat{X},\hat{Y}]_m}{m!}=\hat{Y}+[\hat{X},\hat{Y}]+\frac{1}{2!}[\hat{X},[\hat{X},\hat{Y}]]+\dots,\\
        \frac{d}{dt}e^{\hat{X}}&=e^{\hat{X}(t)}\int_0^1\mathrm{d}\lambda\; e^{-\lambda\hat{X}(t)} \hat{X}'(t) e^{\lambda \hat{X}(t)}
\end{align}
leads to the Hamiltonian in the lattice frame
\begin{align} \label{eq:Hamiltonian} 
        H(t)&=i\hbar \dot{U}(t)U^{\dagger}(t)+U(t)H_{\mathrm{lab\; frame}}(t)U^{\dagger}(t)\notag\\
        &=-\dot{x}_L(t)\hat{p}+\dot{p}_L(t)\hat{x}-\hbar \dot{\Phi}_U(t)+\frac{(\hat{p}-p_L(t))^2}{2m}+V_0(t) \cos^2(k_L\hat{x})\notag\\
        &=\frac{\hat{p}^2}{2m}+V_0(t) \cos^2(k_L \hat{x})+ma_L(t)\hat{x}.
\end{align}
\subsection{Reduced lattice frame}
It will be useful to additionally transform the Hamiltonian~\eqref{eq:HamiltonianLab} to the reduced lattice frame by the unitary transformation
\begin{align} \label{eq:unitaryReducedLattice} 
        \tilde{U}(t)=\exp(i\hat{p}x_L(t)/\hbar+i\Phi_U(t))
\end{align}
with $\dot{\Phi}_U(t)=p_L^2(t)/2m$. This yields 
\begin{align} \label{eq:HamiltonianReducedLattice} 
        H_{\mathrm{reduced}}(t)&=i\hbar \dot{\tilde{U}}(t)\tilde{U}^{\dagger}(t)+\tilde{U}(t)H_{\mathrm{lab\; frame}}(t)\tilde{U}^{\dagger}(t)\notag\\
        &=-\dot{x}_L(t)\hat{p}-\hbar \dot{\Phi}_U(t)+\frac{(\hat{p}-p_L(t))^2}{2m}+V_0(t) \cos^2(k_L\hat{x})\notag\\
        &=\frac{(\hat{p}-p_L(t))^2}{2m}+V_0(t) \cos^2(k_L \hat{x}).
\end{align}

\section{Bloch oscillations in Bloch basis}\label{App:Bloch oscillations in Bloch basis}
In this section, we aim to derive the most common description of Bloch oscillations based on the adiabatic theorem using instantaneous Bloch states~\cite{Bloch1929,Zener1934}. The eigenstates of the acceleration free Hamiltonian $H_0=\hat{p}^2/2m+V_0\cos^2(k_L\hat{x})$ are Bloch states $\ket{\alpha,\kappa}$
\begin{align}\label{eq:H0}  
    H_0\ket{\alpha,\kappa}=E_{\alpha}(\kappa)\ket{\alpha,\kappa},
\end{align}
where $\kappa$ denotes quasi-momentum and $\alpha$ the band number. The eigenstates can be written in a position basis with coefficients $\psi_{\alpha,\kappa}(x)=\bra{x}\ket{\alpha,\kappa}=e^{i\kappa x}u_{\alpha,\kappa}(x)$ and a spatially periodic part $u_{\alpha,\kappa}(x+d)=u_{\alpha,\kappa}(x)$. Inserting $\psi_{\alpha,\kappa}(x)$ into the Schrödinger equation~\eqref{eq:H0} yields
\begin{align}\label{eq:PeriodicBloch}
   %&\left(\frac{(\hat{p}+\hbar \kappa)^2}{2m}+V_0 \cos^2(k_L \hat{x})\right)u_{\alpha,\kappa}(x)=E_{\alpha}(\kappa)u_{\alpha,\kappa}(x)\notag\\
   H(\kappa)u_{\alpha,\kappa}(x)=E_{\alpha}(\kappa)u_{\alpha,\kappa}(x),
\end{align}
where $H(\kappa)=(\hat{p}+\hbar \kappa)^2/2m+V_0 \cos^2(k_L \hat{x})$. We proceed with the time-dependent problem in the reduced lattice frame. In this frame, the Hamiltonian has a discrete translational symmetry, expressed by $[\hat{T}_{d\ell},H_{\mathrm{reduced}}(t)]=0$ for all times $t$, where we defined the discrete translation operator as $\hat{T}_{d\ell}=\exp(i\hat{p}d\ell /\hbar)$. This implies that the quasi-momentum $\kappa \in [-k_L,k_L]$ is conserved under the evolution of $H_{\mathrm{reduced}}(t)$ and there exists a simultaneous eigenbasis of $\hat{T}_{d\ell}$ and $H_{\mathrm{reduced}}(t)$ in the form of
\begin{align}\label{eq:ansatzBloch}
    \psi_{\kappa}(x,t)=e^{i\kappa x}u(x,t)
\end{align}
with a spatially periodic part $u(x+d,t)=u(x,t)$. Inserting $\psi_{\kappa}(x,t)$ into the time-dependent Schrödinger equation and defining the time-dependent quasi-momentum as $\kappa(t)=\kappa-p_L(t)/\hbar$ yields
\begin{align}\label{eq:SchrödingerUXT}
    i\hbar\partial_t u(x,t)=H(\kappa(t))u(x,t),
\end{align}
where $H(\kappa(t))=(\hat{p}+\hbar\kappa(t))^2/2m+V_0 \cos^2(k_L \hat{x})$.  According to Eq.~\eqref{eq:PeriodicBloch} and due to the periodicity of Bloch states in the quasi-momentum $\ket{\alpha,\kappa+2\pi/d}=\ket{\alpha,\kappa}$ the eigenstates of the time-dependent Hamiltonian $H(\kappa(t))$ are given by the periodic part of Bloch states $\ket*{u_{\alpha,\kappa(t)}}$ with a time-dependent quasi-momentum $\kappa(t)$
\begin{align}
    H(\kappa(t))\ket{u_{\alpha,\kappa(t)}}=E_{\alpha}(\kappa(t))\ket{u_{\alpha,\kappa(t)}}.
\end{align}
As described in the main article, we consider atoms localized in the fundamental Bloch band $\alpha=0$ with an atomic momentum width $\varphi(\kappa)$ that vanishes outside the first Brillouin zone $[-k_L,k_L]$. Applying the adiabatic approximation and transforming from the reduced lattice frame to the lattice frame, results in
\begin{align}\label{eq:BlochDynamics}
    \ket{\psi(t)}&=\exp(-i\hat{x}p_L(t)/\hbar)\int\mathrm{d}\kappa\;\varphi(\kappa)e^{i\kappa x}e^{-i\int_0^t\mathrm{d}\tau\;E_0(\kappa(\tau))/\hbar}\,\ket{u_{\alpha,\kappa(t)}}\notag\\
    &=\int\mathrm{d}\kappa\;\varphi(\kappa)e^{-i\int_0^t\mathrm{d}\tau\;E_0(\kappa(\tau))/\hbar}\ket{0,\kappa(t)}.
\end{align}
Eq.~\eqref{eq:BlochDynamics} describes the dynamics of Bloch oscillations of atoms in optical lattices in the lattice frame under the adiabatic approximation performed in the reduced lattice frame. We point out, that due to the time-dependent unitary transformation $\exp(-i\hat{x}p_L(t)/\hbar)$ that connects the reduced lattice frame and the lattice frame, the notion of adiabaticity in both frames is not equivalent. As described in the main article, non-adiabatic corrections in the reduced lattice frame will occur at avoided crossings where atoms can undergo transitions to neighbouring Bloch bands, which is captured by the Landau-Zener formula~\cite{Zener1934}
\begin{align} \label{eq:LZ}
        P_{LZ}=\exp(\frac{-\pi^2\left(\frac{\Delta E}{E_r}\right)^2}{8\frac{dma_L}{E_r}}),
\end{align}
where $E_r=\hbar^2k_L^2/2m$ denotes the recoil energy and $\Delta E$ the energy gap of the fundamental and first excited Bloch bands at the avoided crossing. To compare Eq.~\eqref{eq:LZ} to the Wannier-Stark model (see Fig.~2 (d)  in the main text) we compute the effective linewdith for atoms to tunnel to the first excited Bloch band by solving the equation $e^{-\Gamma_BNT_B/\hbar}=(1-P_{LZ})^N$ resulting in 
\begin{align}\label{eq:LZgamma}
    \Gamma_B=\frac{dma_L}{2\pi}\ln(1-P_{LZ})\simeq\frac{dma_L}{2\pi}P_{LZ}
\end{align}
using the power series expansion of the natural logarithm. 

\section{Wannier-Stark states and crossing scenarios for non-hermitian Hamiltonians}\label{App:} 
In this section, we provide a summary of basic definitions and properties of Wannier-Stark states, essentially following the ideas outlined in~\cite{Gluck2002, Gluck1998}. We assume a constant and non-vanishing acceleration $a_L(t)=a_L>0$ and lattice depth $V_0(t)=V_0>0$ and work in the lattice frame, which gives rise to a time-independent Hamiltonian $H(t)=H$ [see Eq.~\eqref{eq:Hamiltonian}]. Due to the inertial force term $ma_L\hat{x}$, the discrete translational symmetry in the lattice frame is broken
\begin{align}\label{eq:ladderProp}
    [\hat{T}_{d\ell},H]=d\ell ma_L\hat{T}_{d\ell}.
\end{align}
Eq.~\eqref{eq:ladderProp} implies that the discrete translational operator $\hat{T}_{d\ell}$ is a ladder operator for the spectrum of $H$, that is, for a given eigenstate $\ket{\Psi_{\mathcal{E}}}$ of $H$ with energy $\mathcal{E}$, we can construct a new eigenstate $\hat{T}_{d\ell}\ket{\Psi_{\mathcal{E}}}$ with energy $\mathcal{E}+d\ell ma_L$ 
\begin{align}
    H\left(\hat{T}_{d\ell}\ket{\Psi_{\mathcal{E}}}\right) = (\mathcal{E}+d\ell ma_L)\left(\hat{T}_{d\ell}\ket{\Psi_{\mathcal{E}}}\right),
\end{align}
where the set of eigenstates $\{\hat{T}_{d\ell}\ket{\Psi_{\mathcal{E}}}\}_{\ell \in \mathds{Z}}$ is called Wannier-Stark ladder. The time-evolution operator $U(t)=\exp(-iHt/\hbar)$ satisfies 
\begin{align}\label{eq:TimeEvo}
    [\hat{T}_{d\ell},U(t)]=(e^{-id\ell ma_Lt/\hbar}-1)U(t)\hat{T}_{d\ell}.
\end{align}
The right-hand side of Eq.~\eqref{eq:TimeEvo} vanishes for multiples of the Bloch period $T_B=2\hbar k_L/ma_L$, indicating the existence of simultaneous eigenstates of the time-evolution operator over one Bloch period $U(T_B)$, denoted as Floquet-Bloch operator, and the ladder operator $\hat{T}_{d\ell}$. These states are known as Wannier-Bloch states $\ket{\Phi_{\alpha,\kappa}}$ and obey
\begin{align}
    U(T_B)\ket{\Phi_{\alpha,\kappa}}&=e^{-i\mathcal{E}_{\alpha,0}T_B/\hbar}\ket{\Phi_{\alpha,\kappa}},\label{eq:WannierBloch}\\
    \hat{T}_{d\ell}\ket{\Phi_{\alpha,\kappa}}&=e^{i\kappa d\ell}\ket{\Phi_{\alpha,\kappa}},
\end{align}
where $\alpha$ denotes the ladder quantum number, $\kappa$ the quasi-momentum and $\mathcal{E}_{\alpha,0}$ a quasi-energy, which is independent of the quasi-momentum $\kappa$~\cite{Gluck2002, Gluck1998} and only defined modulo $d m a_L$ due to $e^{-i(\mathcal{E}_{\alpha,0}+d\ell ma_L)T_B/\hbar}=e^{-i\mathcal{E}_{\alpha,0}T_B/\hbar}$. Based on the Wannier-Bloch states $\ket*{\Phi_{\alpha,\kappa}}$ we construct Wannier-Stark states $\ket*{\Psi_{\alpha,\ell}}$ with lattice site quantum number $\ell$ according to 
\begin{align} \label{eq:WannierStark}
    \ket{\Psi_{\alpha,\ell}}&=\sqrt{\frac{d}{2\pi}}\int_{-k_L}^{k_L}\mathrm{d}\kappa\, e^{-i\kappa d \ell} \ket{\Phi_{\alpha,\kappa}}.
\end{align}
These states are (quasi-)bound eigenstates of $H$~\cite{Gluck2002, Gluck1998}
\begin{align} \label{eq:WannierStarkEigenvalue}
    H\ket{\Psi_{\alpha,\ell}}=\left(\mathcal{E}_{\alpha,0}+d\ell ma_L\right)\ket{\Psi_{\alpha,\ell}},
\end{align}
where the quasi-energy $\mathcal{E}_{\alpha,0}+d\ell ma_L=\mathcal{E}_{\alpha,\ell}$, as defined in Eq.~\eqref{eq:WannierBloch}, is denoted as Wannier-Stark energy. This shows, that the eigenvalue problem of the Hamiltonian $H$ can be solved by diagonalizing the Floquet-Bloch operator $U(T_B)$ [see Eq.~\eqref{eq:WannierBloch}], providing access to the Wannier-Stark energy $\mathcal{E}_{\alpha,\ell}$ as well as the Wannier-Stark states $\ket*{\Psi_{\alpha,\ell}}$ using Eq.~\eqref{eq:WannierStark}. 

We proceed to present a numerical routine to calculate the complex Wannier-Stark energies $\mathcal{E}_{\alpha,0}$~\cite{Gluck2002, Gluck1998}. As described, the main idea is to solve Eq.~\eqref{eq:WannierBloch}, which is achieved by rewriting the Floquet-Bloch operator in the following form
\begin{align} \label{eq:FloquetBloch}
    U(T_B)=e^{-i2\hat{x}k_L}U_{\mathrm{reduced}}(T_B),
\end{align}
where $U_{\mathrm{reduced}}(T_B)$ is the time-ordered evolution operator in the reduced lattice frame over one Bloch period $T_B$
\begin{align} 
    %U_{\mathrm{reduced}}(T_B)=\mathcal{T}\exp(-\frac{i}{\hbar}\int_0^{T_B}\mathrm{d}t\;\left(\frac{(\hat{p}-ma_Lt)^2}{2m}+V_0 \cos^2(k_L \hat{x})\right))
    U_{\mathrm{reduced}}(T_B)=\mathcal{T}\exp(-\frac{i}{\hbar}\int_0^{T_B}\mathrm{d}t\;H_{\mathrm{reduced}}(t))
\end{align}
and $\mathcal{T}$ denotes time ordering. To compute $U_{\mathrm{reduced}}(T_B)$, we define a time grid with $J$ equidistant grid points $t_j=(j-1/2)dt$ for $j\in \{1,...,J\}$ and a time step of $dt=T_B/J$. For sufficiently small time steps $dt$, we approximate
\begin{align}\label{eq:UreducedTime}
    U_{\mathrm{reduced}}(T_B)\simeq \prod_{j=1}^J\exp(-i H_{\mathrm{reduced}}(t_j)dt/\hbar ).
\end{align}
Next, we expand the Floquet-Bloch operator in a momentum basis $\ket{2n \hbar k_L+\hbar \kappa}$ with $n \in \mathds{Z}$ and quasi-momentum $\kappa\in[-k_L,k_L]$. As described, the Wannier-Stark energies $\mathcal{E}_{\alpha,0}$ are independent of the quasi-momentum $\kappa$ and hence, without loss of generality, we choose $\kappa=0$ and find 
\begin{align} \label{eq:TruncationH}
    \sum_{n \in \mathds{Z}} H_{\mathrm{reduced}}(t_j) \ket{2n\hbar k_L}\bra{2n\hbar k_L}=\sum_{n \in \mathds{Z}}\left(\frac{(2n\hbar k_L-ma_Lt_j)^2}{2m}\ket{2n\hbar k_L}\bra{2n\hbar k_L}+\frac{V_0}{4}\left(\ket{2(n+1)\hbar k_L}\bra{2n\hbar k_L}+\mathrm{H.c.}\right)\right),
\end{align}
where we subtracted the average ac Stark shift $V_0/2$. The momentum shift operator $ e^{-i2\hat{x}k_L}$ takes the following form 
\begin{align} \label{eq:MomentumShift}
    \sum_{n \in \mathds{Z}} e^{-i2\hat{x}k_L} \ket{2n\hbar k_L}\bra{2n\hbar k_L}=\ket{2(n-1)\hbar k_L}\bra{2n\hbar k_L}.
\end{align}
The combination of Eqs.~\eqref{eq:UreducedTime}, \eqref{eq:TruncationH} and \eqref{eq:MomentumShift} yields the discretized Floquet-Bloch operator $U(T_B)$ [see Eq.~\eqref{eq:FloquetBloch}] based on a discretized time grid with $J$ grid points and a discretized momentum grid truncated at $n\in[-N,...,N]$ with $2N+1$ grid points. This leads to a low-dimensional matrix, that can be efficiently constructed and diagonalized using standard libraries for scientific computing, such as SciPy~\cite{Virtanen2020}, while the convergence in the discretization parameters $J$ and $N$ has to be ensured to achieve accurate results. 
%Truncating the momentum shift operator $ e^{-i2\hat{x}k_L}$ leads to a non-unitary matrix, as seen by Eq.~\eqref{eq:MomentumShift}, which results in $2N+1$ complex eigenvalues of the Floquet-Bloch operator $U(T_B)$ in the form $\lambda_{n}=\exp(-i\mathcal{E}_nT_B/\hbar)=\exp(-i(E_n-i\Gamma_n/2)T_B/\hbar)$, as derived in Eq.~\eqref{eq:WannierBloch} and described in the main article in Eq.~(2). 
Truncating the momentum shift operator $e^{-i2\hat{x}k_L}$ leads to a non-unitary matrix, as seen by Eq.~\eqref{eq:MomentumShift}, which results in $2N+1$ complex eigenvalues of the Floquet-Bloch operator $U(T_B)$ in the form $\lambda_{n}=\exp(-i\mathcal{E}_nT_B/\hbar)=\exp(-i(E_n-i\Gamma_n/2)T_B/\hbar)$. This allows to determine the Wannier-Stark energies $\mathcal{E}_n$ modulo $dma_L$. To compute the eigenvalues of $H$, i.e. the Wannier-Stark energies $\mathcal{E}_{\alpha,\ell}$ we need to reconstruct the modulo operation, which additionally implies that the momentum state index $n$ needs to be linked to the correct ladder quantum number $\alpha$. Since the real part of the Wannier-Stark energies $\mathrm{Re}(\mathcal{E}_{\alpha,0})=E_{\alpha,0}$ for typical combinations of lattice depth $V_0$ and acceleration $a_L$ realizing LMT Bloch pulses is larger than the acceleration energy $dma_L$, sorting according to increasing energies $E_n$ will not result in the correct identification of $n$ and $\alpha$. Instead, the Wannier-Stark linewidths $\mathrm{Im}(\mathcal{E}_{\alpha,0})=\Gamma_{\alpha}/2\ll dma_L$ are typically much smaller than the acceleration energy $dma_L$ and hence, sorting according to increasing linewdiths $\Gamma_n$ results in the correct mapping between $n$ and $\alpha$. 
\begin{figure}[t]
    \includegraphics[width=0.5\textwidth,keepaspectratio]{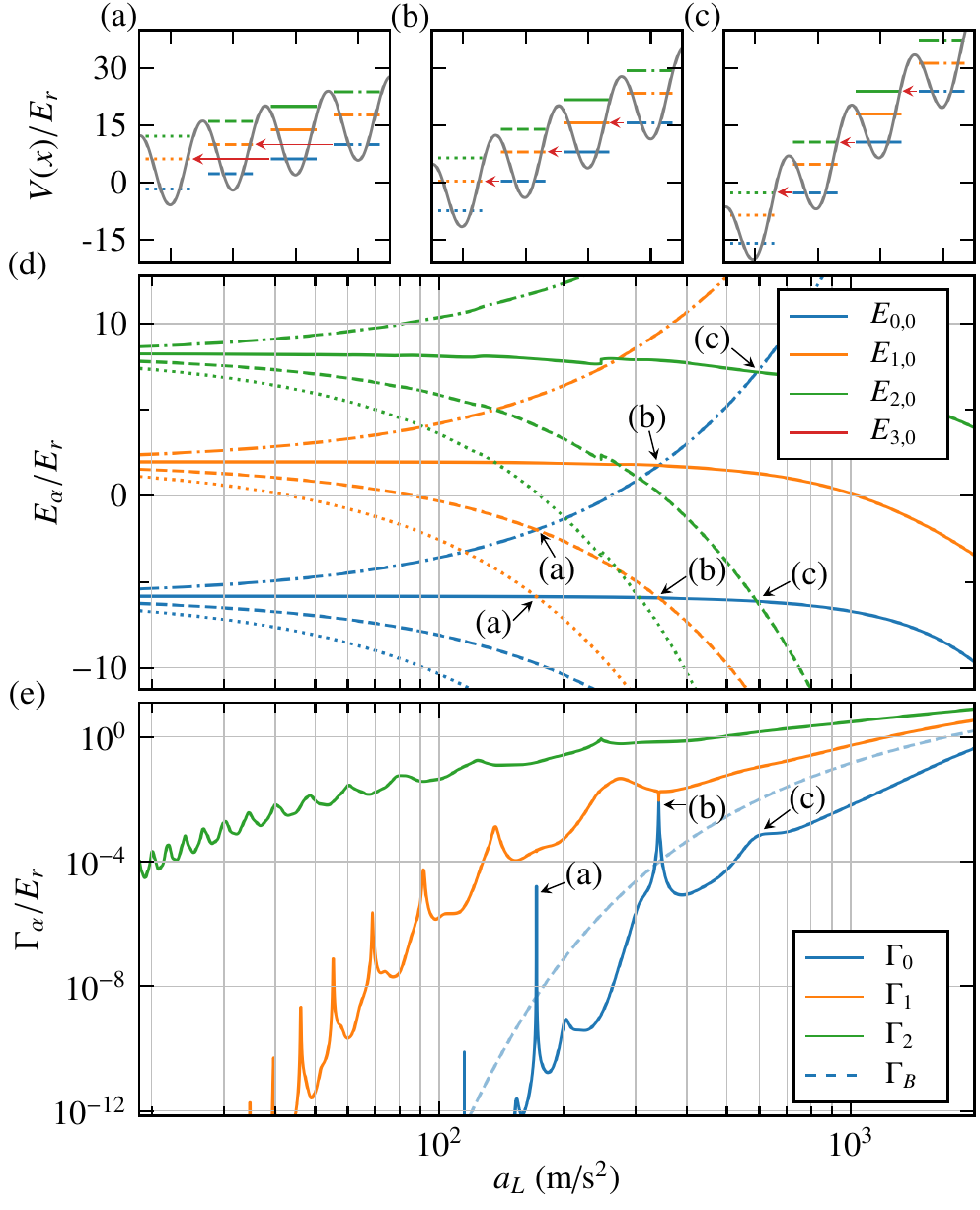}
    \caption{(a)-(c) Tilted potential [see Eq.~\eqref{eq:Hamiltonian}] for specific accelerations indicated in (d,e) and a lattice depth of $V_0=20\,E_r$, including corresponding Wannier-Stark energy levels. Red arrows represent process of resonant tunneling between different Wannier-Stark ladders. (d) Real Wannier-Stark energies for $V_0=20\,E_r$ versus peak acceleration $a_L$ for the first three Wannier-Stark ladders $\alpha=0$ (blue lines), $\alpha=1$ (orange lines) and $\alpha=2$ (green lines). For all ladders the nearest-neighbour energy to the right (dashed-dotted line), the nearest-neighbour energy to the left (dotted line) and the next-nearest-neighbour energy to the left (dotted line) are shown, as indicated in (a)-(c). (e) Linewidth of Wannier-Stark ladders for $V_0=20\,E_r$ versus peak acceleration $a_L$. Solid lines show linewidths for the first three Wannier-Stark ladders $\alpha=0,1,2$, while dashed line shows the linewidth for the fundamental Wannier-Stark ladder predicted by the Landau-Zener formula [see Eq.~\eqref{eq:LZgamma}].}\label{fig:WSspectrumFull}
\end{figure}
This type of sorting, however, will fail for small accelerations, where the computed linewidths $\Gamma_n$ for several Wannier-Stark ladders drop below the precision limit of approximately $10^{-14}\,E_r$ that can be efficiently achieved for standard scientific libraries~\cite{Virtanen2020}. In this regime the Wannier-Stark linewidths are indistinguishable from one another, as seen in Fig.~\ref{fig:WSspectrumFull}(e) for $a_L\lesssim
150\,\mathrm{m}/\mathrm{s}^2$. We develop a sorting algorithm based on an approximate formula for the real Wannier-Stark energies $E_{\alpha,\ell}$~\cite{Gluck2002, Gluck1998} valid for small accelerations $a_L \ll V_0/dm$, given by
\begin{align} \label{eq:WSapprox}
    E_{\alpha,\ell}(V_0,a_L)&\simeq\langle E^{Bloch}_{\alpha}(V_0) \rangle+E_{\Delta x}(V_0,a_L)+dlm a_L,\\
    E_{\Delta x}(V_0,a_L)&=\frac{V_0}{2}\sin(-\arcsin(\frac{ma_L}{k_L V_0}))^2-ma_L\arcsin(\frac{ma_L}{k_L V_0}), \notag
\end{align}
where $\langle E^{Bloch}_{\alpha}(V_0)\rangle$ denotes the average energy of the $\alpha$th Bloch band. We sort the Wannier-Stark energies by identifying each numerical solution with the closest approximate solution given by Eq.~\eqref{eq:WSapprox}. For very small accelerations the Bloch period starts to diverge $T_B\propto 1/a_L$ and the numerical construction of the Floquet-Bloch operator becomes inefficient. In this region, we simply rely on the approximate solution of the real Wannier-Stark energies $E_{\alpha,\ell}(V_0,a_L)$. The generalizations presented, enable the treatment of adiabatic LMT Bloch pulses through numerical diagonalization of the Floquet-Bloch operator and the accurate computation of the evolved phase $\phi=e^{-i\int_0^t\mathrm{d}t'E_{0,0}(t')/\hbar}$ and tunneling losses  $1-P=1-e^{-\int_0^t\mathrm{d}t'\Gamma_{0}(t')/\hbar}$, as shown in the main article.

\begin{figure}[t]
        \includegraphics[width=1\textwidth,keepaspectratio]{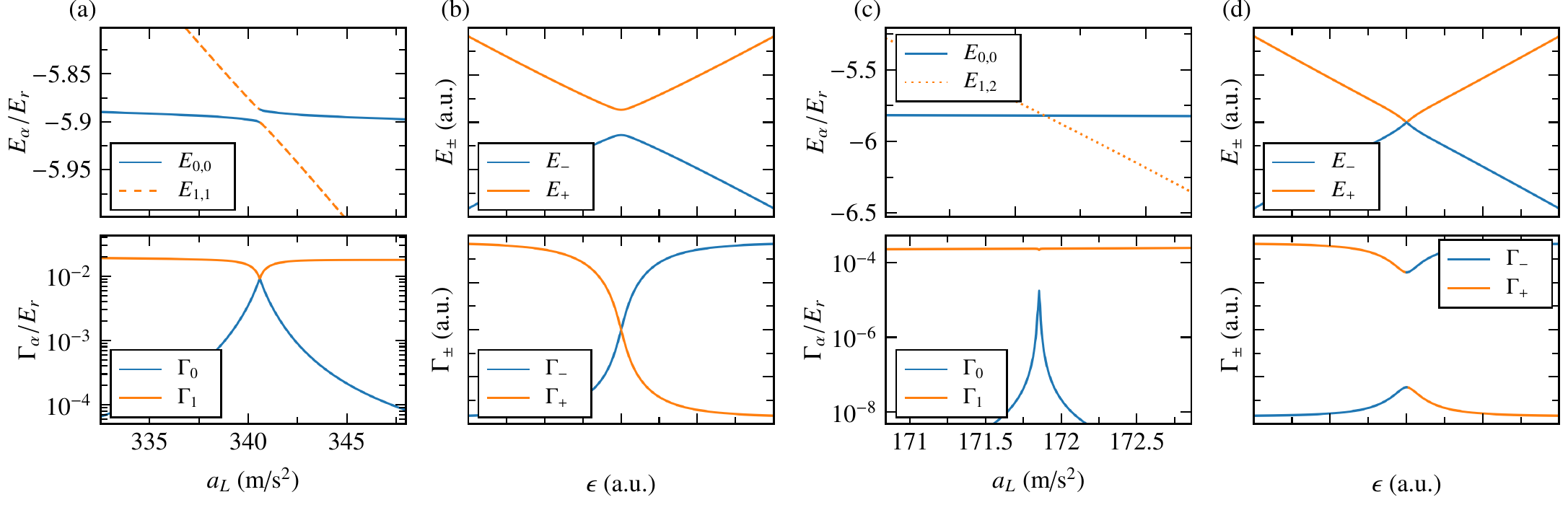}
        \caption{(a) Magnified view of Fig.~\ref{fig:WSspectrumFull}(d,e) displaying a crossing type $I$. (b) Real $E_{\pm}$ and imaginary $\Gamma_{\pm}$ part of Eq.~\eqref{eq:LZmodel} for $|V|>\gamma$, qualitatively capturing the behaviour of a type $I$ crossing. (c) Magnified view of Fig.~\ref{fig:WSspectrumFull}(d,e) displaying a crossing type $II$. (d) Real $E_{\pm}$ and imaginary $\Gamma_{\pm}$ part of Eq.~\eqref{eq:LZmodel} for $|V|<\gamma$, qualitatively capturing the behaviour of a type $II$ crossing.}\label{fig:Crossings}
\end{figure}

Fig.~\ref{fig:WSspectrumFull} shows an extension of Fig.~2 of the main article, additionally displaying real Wannier-Stark energies $E_{\alpha,\ell}$, including the nearest and next-nearest-neighbour energy levels. It is clearly visible, that each tunneling resonance, as seen in Fig.~\ref{fig:WSspectrumFull}(e) corresponds to a real energy crossing displayed in Fig.~\ref{fig:WSspectrumFull}(d), as described in the main article. The primary tunneling resonances of the fundamental Wannier-Stark ladder occur when energies of nearest or next-nearest neighbors cross, since atoms experience larger tunneling strengths in these cases compared to energy crossings involving more distant neighbors. Indeed, we can distinguish two different crossing types that appear for typical Wannier-Stark spectra relevant for LMT Bloch pulses, as shown in Figs.~\ref{fig:Crossings}(a) and \ref{fig:Crossings}(c). In the first crossing scenario, denoted as type $I$, the real Wannier-Stark energies $E_{\alpha,\ell}$ form an avoided crossing, while the Wannier-Stark linewidths $\Gamma_{\alpha}$ exhibit a real crossing, as shown in Fig.~\ref{fig:Crossings}(a). In the second crossing scenario, denoted as type $II$, the Wannier-Stark linewidths $\Gamma_{\alpha}$ form an avoided crossing, while the real Wannier-Stark energies $E_{\alpha,\ell}$ exhibit a real crossing, as shown in Fig.~\ref{fig:Crossings}(c). The occurrence of different crossing types of the Wannier-Stark spectrum can be explained with a simple model, based on a two-level system~\cite{Mondragon1996, Keck2003, Ibanez2014}. We assume an energy level splitting of $\Delta E=2\epsilon$, where the excited level has a much larger linewidth than the ground state $\gamma=\gamma_1\gg\gamma_0\approx 0$ and a coupling strength between the two bare states of $V$. The corresponding non-hermitian Hamiltonian reads

\begin{align}
H=
    \begin{pmatrix}
        \epsilon-2i\gamma & V \\
        V & -\epsilon 
    \end{pmatrix}
\end{align}
with complex eigenvalues given by
\begin{align}\label{eq:LZmodel}
    \mathcal{E}_{\pm}=-i\gamma \pm \sqrt{(\epsilon-i\gamma)^2+V^2}=E_{\pm}-i\Gamma_{\pm}/2.
\end{align}
As shown in Figs.~\ref{fig:Crossings}(b) and \ref{fig:Crossings}(d) this model predicts the behaviour of type $I$ and type $II$ crossings for $|V|>\gamma$ and $|V|<\gamma$, respectively. This shows, that the ratio of the coupling strength between the two bare states and the linewidth of the excited state determines the crossing type.

\section{Spontaneous emission losses}\label{App:}
In this section, we treat an additional important loss channel for LMT Bloch pulses, given by spontaneous emission losses. We adopt a simplified perspective, where atoms lost due to a spontaneous scattering event are considered to only reduce the overall atom number. However, in reality, these lost atoms can undergo transitions to different Wannier-Stark ladders of the optical lattice, resulting in more complex dynamics beyond simple particle loss. Nevertheless, for sufficiently small spontaneous emission losses, these effects can be effectively suppressed. We estimate the spontaneous scattering rate using~\cite{Grimm2000}
\begin{align}\label{eq:SpontEmAverage}
    \hbar\Gamma_{\mathrm{sp}}&=\frac{\Gamma_{\mathrm{nat}}}{|\Delta|}\langle V(\hat{x}) \rangle,
\end{align}
where $\langle V(\hat{x}) \rangle$ denotes the average value of the potential w.r.t. to the atomic state, $\Gamma_{\mathrm{nat}}$ the natural line width of the atomic transition and $\Delta$ the detuning from the atomic resonance frequency. To reduce the influence of spontaneous emission, it is useful to work with blue-detuned optical lattices $\Delta>0$, as commonly done in state-of-the-art experiments~\cite{Gebbe2021, Morel2020}, since atoms will be localized in low-intensity regions of the optical lattice, which in turn reduces the spontaneous scattering rate $\Gamma_{\mathrm{spont.}}$ [see Eq.~\eqref{eq:SpontEmAverage}]. For deep optical lattices ($V_0\gg E_r$) using the harmonic approximation we find
\begin{align}
    \hbar\Gamma_{\mathrm{sp}}&=\frac{\Gamma_{\mathrm{nat}}}{|\Delta|}\frac{\sqrt{V_0E_r}}{2}.
\end{align}
The resulting surviving fraction of atoms is given by
\begin{align}\label{eq:lossesSE}
    P_{\mathrm{sp}}=\exp(-\Gamma_{\mathrm{sp}}T),
\end{align}
where $T$ is the acceleration time, as defined in the main article. The lattice depth $V_0$ is related to the power $P$ and waist $w$ of a Gaussian laser beam via~\cite{Grimm2000}
\begin{align}
    V_0=\frac{3\pi c^2}{2\omega_0^3}\frac{\Gamma_{\mathrm{nat}}}{|\Delta|} \frac{2 P}{\pi w^2},
\end{align}
where $c$ denotes the speed of light in vaccuum and $\omega_0$ the atomic resonance frequency. This results in 
\begin{align}\label{eq:LossesSEfinal}
    \hbar\Gamma_{\mathrm{sp}}&=\frac{2\omega_0^3}{3\pi c^2}\frac{\sqrt{V_0^3E_r}}{2I_0},
\end{align}
where $I_0=2P/\pi w^2$ denotes the intensity of the laser beam, showing that for a given laser power $P$ and waist $w$ spontaneous emission losses $1-P_{\mathrm{sp}}$ increase for larger lattice depths $V_0$. 
\begin{figure}[t]
        \includegraphics[width=0.5\textwidth,keepaspectratio]{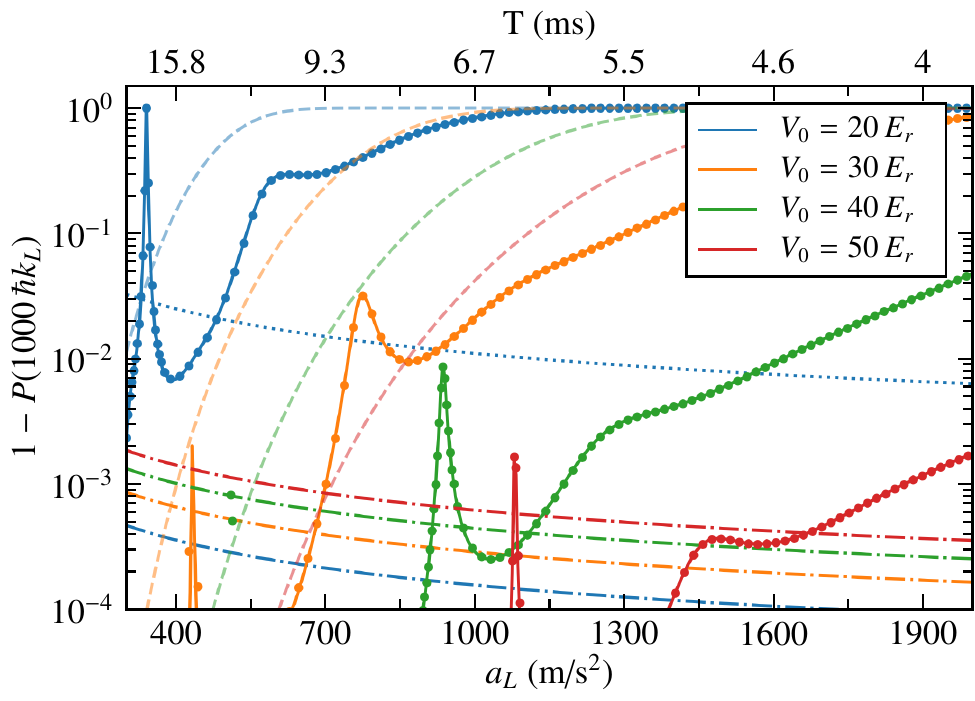}
        \caption{Losses for an adiabatic $1000\,\hbar k_L$ LMT Bloch pulse versus peak acceleration $a_L$ for a given lattice depth $V_0$. The upper axis shows the corresponding acceleration time $T$. Solid lines represent the predicted losses based on the Wannier-Stark model [see Eq.~(4) in the main article], while dots show the exact numerical solution. Dashed lines show the predicted losses based on the Landau-Zener formula [see Eq.~\eqref{eq:LZ}] and dotted lines show the spontaneous emission losses based on a state-of-the-art laser system from Gebbe \textit{et al}.~\cite{Gebbe2021}, while dashed-dotted lines show spontaneous emission losses based on an improved laser system from Kim \textit{et al}.~\cite{Kim2020}.}\label{fig:spontaneousEmission}
\end{figure}
In Fig.~\ref{fig:spontaneousEmission} we compare tunneling losses, as described in the main article, with spontaneous emission losses using Eqs.~\eqref{eq:LossesSEfinal} and \eqref{eq:lossesSE} based on a state-of-the-art laser system from Gebbe \textit{et al}.~\cite{Gebbe2021} with a power of $P=1.2\,\mathrm{W}$ and a waist of $w=3.75\,\mathrm{mm}$ and an improved laser system from Kim \textit{et al}.~\cite{Kim2020} with $P=6\,\mathrm{W}$ and $w=1\,\mathrm{mm}$. Both laser system work with $^{87}\mathrm{Rb}$ addressing the D2 transition. All curves show that for larger accelerations spontaneous emission losses reduce with shorter acceleration times $T$, as evident from Eq.~\eqref{eq:lossesSE}. At the optimal peak acceleration for $V_0=20\,E_r$ at $a_L=393.5$ $\mathrm{m/s}^2$ the state-of-the-art laser system gives rise to three times larger spontaneous emission losses than tunneling losses. We stress that spontaneous emission losses can be reduced when operating with a more powerful laser system or by the possibility of working with a smaller beam waist, as shown in Fig.~\ref{fig:spontaneousEmission}, where the spontaneous emission losses for the improved laser system is more than one order of magnitude smaller than tunneling losses. For larger lattice depths $V_0>20\,E_r$ and assuming maximal tunneling losses of $1\%$, spontaneous emission losses are orders of magnitude smaller than tunneling losses when working with the improved laser system. 

\section{Non-adiabatic losses}\label{App:}
\begin{figure}[t]
        \includegraphics[width=1\textwidth,keepaspectratio]{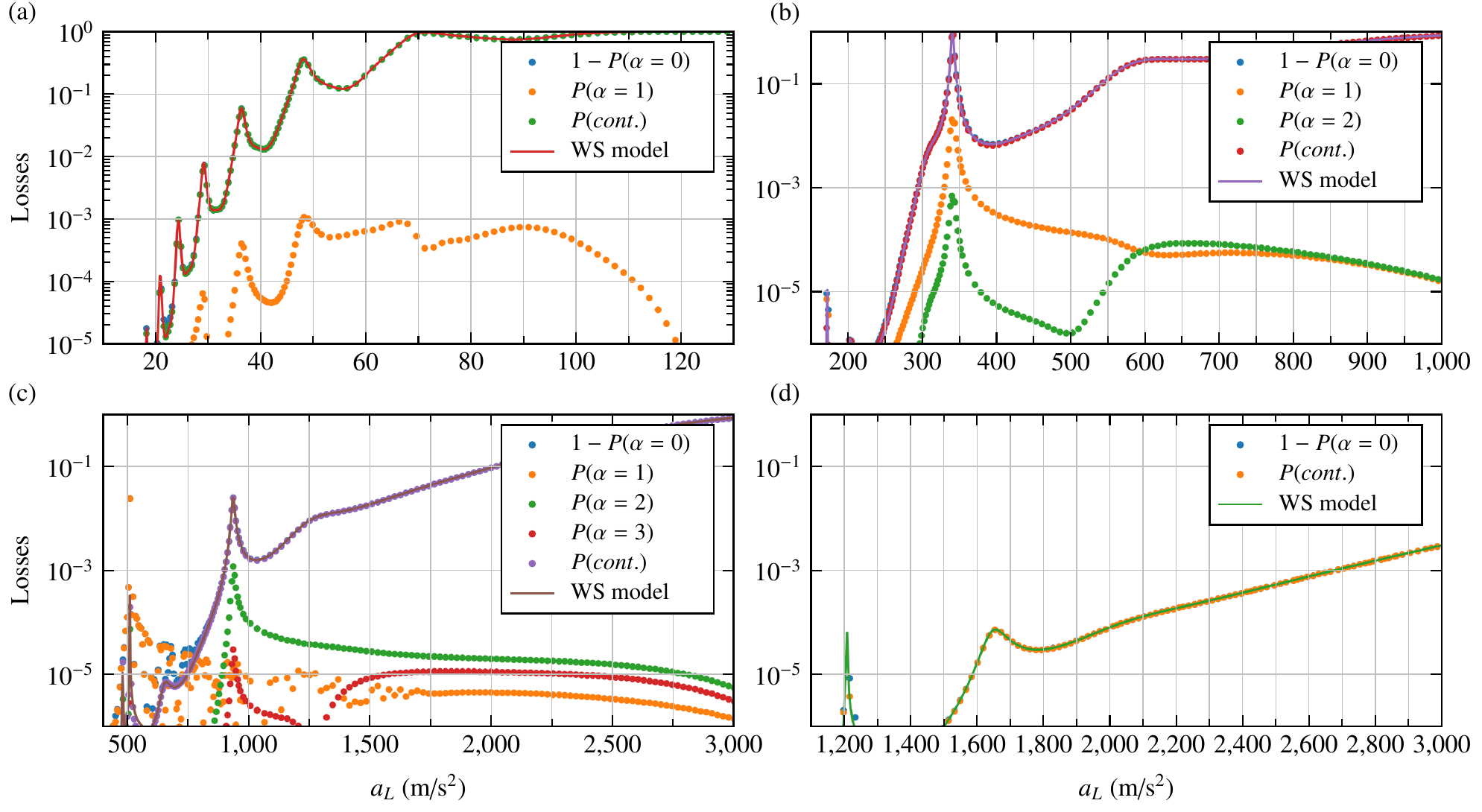}
        \caption{(a-d) Losses for adiabatic LMT Bloch pulses versus peak acceleration $a_L$. (a) $100\,\hbar k_L$ LMT Bloch pulse with a peak lattice depth of $V_0=5\,E_r$. Red solid line represents the predicted losses based on the Wannier-Stark model, while dots show the exact numerical solution, distinguishing between total losses from the Wannier-Stark ladder $\alpha=0$ (blue dots), losses to the Wannier-Stark ladder $\alpha=1$ (orange dots) and tunneling losses (green dots). (b) $1000\,\hbar k_L$ LMT Bloch pulse with a peak lattice depth of $V_0=20\,E_r$. Purple solid line represents the predicted losses based on the Wannier-Stark model, while dots show the exact numerical solution, distinguishing between total losses from the Wannier-Stark ladder $\alpha=0$ (blue dots), losses to the Wannier-Stark ladder $\alpha=1$ (orange dots), losses to the Wannier-Stark ladder $\alpha=2$ (green dots) and tunneling losses (red dots). (c) $1000\,\hbar k_L$ LMT Bloch pulse with a peak lattice depth of $V_0=40\,E_r$. Brown solid line represents the predicted losses based on the Wannier-Stark model, while dots show the exact numerical solution, distinguishing between total losses from the Wannier-Stark ladder $\alpha=0$ (blue dots), losses to the Wannier-Stark ladder $\alpha=1$ (orange dots), losses to the Wannier-Stark ladder $\alpha=2$ (green dots), losses to the Wannier-Stark ladder $\alpha=3$ (red dots) and tunneling losses (purple dots). (d) $1000\,\hbar k_L$ LMT Bloch pulse with a peak lattice depth of $V_0=60\,E_r$. The green solid line represents the predicted losses based on the Wannier-Stark model, while dots show the exact numerical solution, distinguishing between total losses from the Wannier-Stark ladder $\alpha=0$ (blue dots) and tunneling losses (orange dots). Losses to excited Wannier-Stark ladders $\alpha>0$ are smaller than $10^{-6}$ and therefore not visible in the depicted regions.}\label{fig:NonAdiabaticLossesCombined}
\end{figure}
As shown in the main article, the fundamental and dominant loss mechanism for adiabatic LMT Bloch pulses is given by tunneling losses. In Fig.~4 of the main article, we observe that non-adiabatic losses of an adiabatic LMT Bloch pulse with an optimal peak acceleration of $a_L=393.5$ $\mathrm{m/s}^2$ and peak lattice depth of $V_0=20\;E_r$ are suppressed by more than one order of magnitude compared to tunneling losses. In Fig.~\ref{fig:NonAdiabaticLossesCombined} we study non-adiabatic losses for adiabatic LMT Bloch pulses in dependency of the peak acceleration $a_L$ for various peak lattice depths ranging from $V_0=5\,E_r$ to $V_0=60\,E_r$ and observe the same behaviour. This confirms that the Wannier-Stark model provides an accurate description for adiabatic LMT Bloch pulses across a wide range of peak lattice depths and accelerations. Only for very small peak accelerations $a_L$, as shown for instance in Fig.~\ref{fig:NonAdiabaticLossesCombined}(c) for $a_L<700$ m/s$^2$, non-adiabatic losses can surpass tunneling losses. These regions, however, are characterized by longer acceleration times $T$, resulting in increased spontaneous emission losses and consequently, are suboptimal for implementing LMT Bloch pulses. 

Fig.~\ref{fig:NonAdiabaticLossesCombined}(b) shows a characteristic loss landscape for non-adiabatic losses of the first and second excited Wannier-Stark ladders, that can be explained by identifying relevant crossings between Wannier-Stark ladders. As described in the main article, by adiabatically tuning the acceleration $a_L(t)$ [see Fig.~1(d) of the main article] atoms move through several crossings, as shown in Fig.~\ref{fig:WSspectrumFull}, resulting in non-adiabatic losses to excited Wannier-Stark ladders. At a peak acceleration of $a_L=340.5$ $\mathrm{m/s}^2$, we observe maximal non-adiabatic losses for both Wannier-Stark ladders $\alpha=1$ and $\alpha=2$. For the first excited Wannier-Stark ladder this can be attributed to the occurrence of a nearest-neighbour type $I$ crossing of the fundamental and the first excited Wannier-Stark ladders at $a_L=340.5$ $\mathrm{m/s}^2$ [crossing of solid  blue and dashed yellow lines in Fig.~\ref{fig:WSspectrumFull}(d)]. The second excited Wannier-Stark ladder will be mainly populated by two crossings that can be seen in Figs.~\ref{fig:WSspectrumFull}(d) at $a_L=305$ $\mathrm{m/s}^2$ [crossing of solid blue and dotted green lines] and $a_L=271.5$ $\mathrm{m/s}^2$ [crossing of solid yellow and dashed green lines]. The first relevant crossing at $a_L=305$ $\mathrm{m/s}^2$ is given by a next-nearest-neighbour type $II$ crossing of the fundamental and the second excited Wannier-Stark ladders, that will excite atoms during the sigmoidal rise and fall of $a_L(t)$ to the Wannier-Stark ladder $\alpha=2$. The second relevant crossing at $a_L=271.5$ $\mathrm{m/s}^2$ is given by a nearest-neighbour type $II$ crossing between the first and the second excited Wannier-Stark ladders. Atoms that were excited to the Wannier-Stark ladder $\alpha=1$, as explained above, can be excited to the Wannier-Stark ladder $\alpha=2$ during the sigmoidal fall of $a_L(t)$. For increasing peak accelerations in the region of $a_L \in [340.5,500]$ $\mathrm{m/s}^2$ there exists no strong ladder-ladder interaction which leads to a reduction of non-adiabatic losses. At $a_L = 589$ $\mathrm{m/s}^2$, however, we observe that the second excited and fundamental Wannier-Stark ladders form a nearest-neighbour type $II$ crossing [see crossing of solid  blue and dashed green lines in Fig.~\ref{fig:WSspectrumFull}(d)], resulting in a larger fraction of atoms in the second excited Wannier-Stark ladder starting at $a_L=500$ $\mathrm{m/s}^2$, as seen in Fig.~\ref{fig:NonAdiabaticLossesCombined}(b). The analysis presented demonstrates that the Wannier-Stark basis offers a comprehensive understanding of the physics of adiabatic LMT Bloch pulses, extending beyond the adiabatic approximation. This hold true for a wide range of peak lattice depths and accelerations, as seen in Figs.~\ref{fig:NonAdiabaticLossesCombined}(a-d).

\section{Lattice-shift method by Clad\'{e} et al.}\label{App:}
\begin{figure}[t]
        \includegraphics[width=1\textwidth,keepaspectratio]{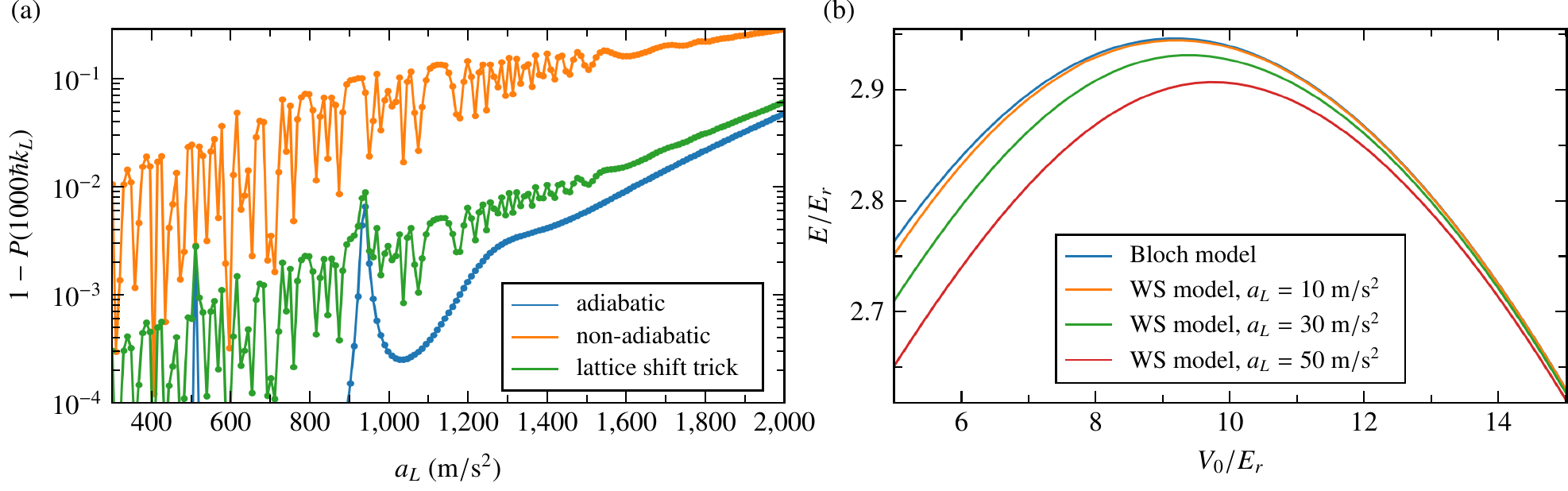}
        \caption{(a) Losses for an $1000\,\hbar k_L$ LMT Bloch pulse versus peak acceleration $a_L$ for a given lattice depth of $V_0=40\,E_r$, including tunneling and non-adiabatic losses. All dots show an exact numerical solution, while the lines are a guide to the eye. Blue dots show losses for an adiabatic LMT Bloch pulse with acceleration ramp time $\tau_{\mathrm{ramp}}=1$ ms, the orange dots show losses for a non-adiabatic LMT Bloch pulse with a vanishing acceleration ramp time $\tau_{\mathrm{ramp}}$, realizing a box acceleration pulse and green dots show losses using the lattice-shift trick as presented in Clad\'{e} \textit{et al.}~\cite{Clade2017} (b) Wannier-Stark energies of the first excited Wannier-Stark ladder $E_{1,0}$ versus peak lattice depth $V_0$ for fixed peak accelerations $a_L$. Blue line is the result of the Bloch model given by Eq.~\eqref{eq:BlochDynamics} and coincides with the result of the Wannier-Stark model based on Eq.~\eqref{eq:WSapprox} in the limit for vanishing peak accelerations $a_L$.}\label{fig:CladeMcAlpineCombined}
\end{figure}
As demonstrated in the main article, the adiabatic control of atoms in Wannier-Stark eigenstates guarantees the achievement of LMT Bloch pulses that reach their fundamental efficiency limit. An acceleration box pulse giving rise to a non-adiabatic LMT Bloch pulse, as analyzed in Fig.~4 of the main article, leads to large non-adiabatic losses to excited Wannier-Stark ladders. The sudden jump in the acceleration $a_L(t)$ results in a sudden shift of the minima of the tilted optical lattice potential $V(x)=V_0 \cos^2(k_L x)+ma_Lx$ given by 
\begin{align}
    \Delta x_{\mathrm{shift}} = -\frac{\arcsin(\frac{ma_L}{k_LV_0})}{2k_L},    
\end{align}
resulting in an oscillatory motion of atoms around the lattice minima. To mitigate this motion, we apply a position shift of $-\Delta x_{\mathrm{shift}}$ to maintain the lattice minima at their original positions, as proposed by Clad\'{e} \textit{et al.}~\cite{Clade2017}. This shift can be applied with a phase shift of one of the lasers that create the optical lattice. Fig.~\ref{fig:CladeMcAlpineCombined} shows that employing the lattice-shift trick significantly reduces the losses of LMT Bloch pulses by over one order of magnitude compared to non-adiabatic LMT Bloch pulses. However, it is important to note that the fundamental efficiency limit can only be attained by adiabatic LMT Bloch pulses, as shown in Fig.~\ref{fig:CladeMcAlpineCombined}. We emphasize that the pronounced oscillations in the losses observed in both the non-adiabatic LMT Bloch pulse with and without the lattice-shift trick serve as evidence for significantly larger non-adiabatic losses compared to tunneling losses. Therefore, in such scenarios, the phase evolution derived in Eq.~(4) of the main article and the analysis of phase uncertainties based on Eq.~(5) of the main article are no longer applicable. 

\section{Magic-lattice-depth Bloch oscillations by McAlpine et al.}\label{App:}
As discussed in the main article, driving adiabatic LMT Bloch pulses in the fundamental Wannier-Stark ladder, characterized by a narrower linewidth compared to excited Wannier-Stark ladders, leads to minimal losses. Nonetheless, it is also feasible to drive LMT Bloch pulses in excited Wannier-Stark ladders $\alpha>0$. This can be accomplished by following the procedure outlined in the main article to drive LMT Bloch pulses, with the adjustment of initially loading atoms in the $\alpha$th Bloch band. The linewidths of excited Wannier-Stark ladders $\Gamma_{\alpha}(t)$ are generally much larger than those of the fundamental Wannier-Stark ladder $\Gamma_0(t)$, as seen in Fig.~\ref{fig:WSspectrumFull}(d). Consequently, the maximal momentum transfer attainable for LMT Bloch pulses in excited Wannier-Stark ladders is significantly constrained. Nevertheless, there are potential advantages in driving such pulses, as highlighted in the work of McAlpine \textit{et al.}~\cite{McAlpine2020}. Based on the common description of Bloch oscillations in a Bloch basis [see section~\ref{App:Bloch oscillations in Bloch basis}] the authors observe that the average Bloch band energy, which determines the evolved phase $\phi=\langle E^{Bloch}_{\alpha}(V_0) \rangle T_B/\hbar$ for one Bloch oscillation (see Eq.~\eqref{eq:BlochDynamics}), exhibits a maximum at $V_0=9.1\,E_r$, as seen in Fig.~\ref{fig:CladeMcAlpineCombined}(b). This means that LMT Bloch pulses driven in the first excited Bloch band at a maximum lattice depth of $V_0=9.1\,E_r$ are insensitive to lattice depth variations to first order. We generalize this result with the help of the Wannier-Stark model. In the limit of vanishing accelerations the Wannier-Stark energies are given by the average Bloch band energy $E_{\alpha,0}(V_0)\approx\langle E^{Bloch}_{\alpha}(V_0)\rangle$ [see Eq.~\eqref{eq:WSapprox}], consistent with the Bloch model. However, for non-vanishing peak accelerations $a_L$ the Wannier-Stark energies change with the peak acceleration $a_L$, leading to a shift in the location of the maximal Wannier-Stark energy $E_{\alpha,0}(V_0,a_L)$, as shown in Fig.~\ref{fig:CladeMcAlpineCombined}(b). This dependency is not covered by the description based on Bloch states. Hence, an accurate determination of the magic lattice depth necessitates the use of the Wannier-Stark model.

%This can be achieved by adiabatically loading atoms in the $\alpha$th Bloch band, corresponding to the $\alpha$th Wannier-Stark ladder, as evident from $\ket{\Psi_{\alpha,\ell}}|_{a_L=0}=\ket{w_{\alpha,\ell}}$, where $\ket{w_{\alpha,\ell}}$ are Wannier states of the Bloch band $\alpha$ localized at lattice site $\ell$. 

%%%%%%%%%%%%%%%%%%%%%%%%%%%%%%%%%%%%%%%%%%%%%%%%%%%%%%%%%%%%%%%%
%%%     Bibliography
%%%%%%%%%%%%%%%%%%%%%%%%%%%%%%%%%%%%%%%%%%%%%%%%%%%%%%%%%%%%%%%% 
\bibliographystyle{apsrev4-1_custom}
\bibliography{my_library}